\documentclass[aps,amsmath,amssymb,twocolumn,nofootinbib,pra,superscriptaddress]{revtex4-1}

\usepackage{graphicx}
\usepackage{bm}
\usepackage{epsfig,amsthm}
\usepackage{color,fancybox}
\usepackage{accents}
\usepackage{times,mathtools}
\usepackage{bbold}
\usepackage[T1]{fontenc}
\usepackage{booktabs}

\usepackage[hyperindex,breaklinks,colorlinks=true,citecolor=blue,urlcolor=blue]{hyperref}

\hypersetup{
	colorlinks=true,
	linkcolor=CitingColor,
	citecolor=CitingColor,
	urlcolor=CitingColor
}

\renewcommand{\url}[1]{\textnormal{#1}}

\newcommand{\ket}[1]{\ensuremath{|#1\rangle}}
\newcommand{\bra}[1]{\ensuremath{\langle #1|}}

\newcommand{\proj}[1]{\ket{#1}\!\bra{#1}}

\newcommand{\vecP}{\vec{P}}

\newcommand{\tA}{\text{A}}
\newcommand{\tB}{\text{B}}
\newcommand{\tC}{\text{C}}

\newcommand{\PABC}{P_\text{ABC}}
\newcommand{\PAB}{P_\text{AB}}

\newcommand{\PAC}{P_\text{AC}}

\newcommand{\vecPABC}{\vecP_\text{ABC}}
\newcommand{\vecPAB}{\vecP_\text{AB}}
\newcommand{\vecPBC}{\vecP_\text{BC}}
\newcommand{\vecPAC}{\vecP_\text{AC}}
\newcommand{\vecPCA}{\vecP_\text{CA}}

\newcommand{\vecPA}{\vecP_\tA}
\newcommand{\vecPB}{\vecP_\tB}

\newcommand{\rAB}{\rho_\text{AB}}
\newcommand{\rBC}{\rho_\text{BC}}
\newcommand{\rAC}{\rho_\text{AC}}
\newcommand{\rABC}{\rho_\text{ABC}}
\newcommand{\vrABC}{\varrho_\text{ABC}}
\newcommand{\vrAC}{\varrho_\text{AC}}

\newcommand{\tr}{{\rm tr}}
\newcommand{\id}{\mathbb{I}}

\newcommand{\C}{\mathcal{C}}
\newcommand{\D}{\mathcal{D}}
\newcommand{\Q}{\mathcal{Q}}
\newcommand{\NS}{\mathcal{NS}}
\newcommand{\cL}{\mathcal{L}}
\newcommand{\CABBC}{\C}
\newcommand{\CsABBC}[1]{\CABBC^#1}

\newtheorem{theorem}{Theorem}
\newtheorem{corollary}[theorem]{Corollary}
\newtheorem{lemma}[theorem]{Lemma}
\newtheorem{definition}{Definition}

\definecolor{CitingColor}{rgb}{0,0.3,1}

\usepackage[capitalize]{cleveref}

\begin{document}
\title{Nonlocality of Quantum States Can be Transitive}

\author{Kai-Siang Chen}
\affiliation{Department of Physics and Center for Quantum Frontiers of Research \& Technology (QFort), National Cheng Kung University, Tainan 701, Taiwan}

\author{Gelo Noel M. Tabia}
\affiliation{Hon Hai (Foxconn) Research Institute, Taipei, Taiwan}
\affiliation{Department of Physics and Center for Quantum Frontiers of Research \& Technology (QFort), National Cheng Kung University, Tainan 701, Taiwan}
\affiliation{Physics Division, National Center for Theoretical Sciences, Taipei 106319, Taiwan}

\author{Chung-Yun Hsieh}
\email{chung-yun.hsieh@bristol.ac.uk}
\affiliation{H. H. Wills Physics Laboratory, University of Bristol, Tyndall Avenue, Bristol, BS8 1TL, United Kingdom}
\affiliation{ICFO - Institut de Ci\`encies Fot\`oniques, The Barcelona Institute of Science and Technology, 08860 Castelldefels, Spain}

\author{Yu-Chun Yin}
\affiliation{Institute of Communications Engineering, National Yang Ming Chiao Tung University, Hsinchu 300093, Taiwan}
\affiliation{Department of Physics and Center for Quantum Frontiers of Research \& Technology (QFort), National Cheng Kung University, Tainan 701, Taiwan}

\author{Yeong-Cherng Liang}
\email{ycliang@mail.ncku.edu.tw}
\affiliation{Department of Physics and Center for Quantum Frontiers of Research \& Technology (QFort), National Cheng Kung University, Tainan 701, Taiwan}
\affiliation{Physics Division, National Center for Theoretical Sciences, Taipei 106319, Taiwan}
\affiliation{Perimeter Institute for Theoretical Physics, Waterloo, Ontario, Canada, N2L 2Y5}

\date{\today}

\begin{abstract}
As a striking manifestation of quantum entanglement, nonlocality has long played a pivotal role in shaping our understanding of the quantum world. When considering a Bell test involving three parties, we may even find a remarkable situation where the nonlocality in two bipartite subsystems {\em forces} the remaining bipartite subsystem to exhibit nonlocality. This intriguing effect, dubbed {\em nonlocality transitivity}, was first identified in the {\em non}-quantum non-signaling world in 2011. However, whether such transitivity could manifest within quantum theory has remained unresolved---until now. Here, we provide the first affirmative answer to this open problem at the level of quantum states, thereby showing that there exists a {\em quantum-realizable} notion of nonlocality transitivity. Specifically, by leveraging the possibility of Bell-inequality violation by tensoring, we analytically construct a pair of nonlocal bipartite states such that simultaneously realizing them in a tripartite system {\em induces} nonlocality in the remaining bipartite subsystem. En route to showing this, we also prove that multiple copies of the $W$-state marginals {\em uniquely} determine the global compatible state, thus establishing another instance when the parts determine the whole. Surprisingly, the nonlocality transitivity of quantum states also occurs among the reduced states of {\em Haar-random} three-qutrit pure states. We further show that the transitivity of {\em quantum steering} can already be demonstrated with the marginals of a three-qubit $W$ state, showing again another noteworthy difference between the two forms of quantum correlations. Finally, we present a simple method to construct quantum states and correlations that are nonlocal in {\em all} their non-unipartite marginals, which may be of independent interest. 
\end{abstract}

\maketitle

\section{Introduction}

Among the various phenomena presented by quantum theory, there is little dispute that quantum entanglement~\cite{Ent-RMP} and the ensuing Bell-nonlocality~\cite{Bell-RMP} stand out as the ones that pose the greatest challenge to our understanding of the physical world. Loosely, the former refers to the profound connection between particles, such that the state of one particle may be strongly or even perfectly correlated to the state of another, regardless of the distance separating them. This ``spooky action at a distance,'' as Einstein~\cite{Einstein1971TheBL} described it, underpins quantum nonlocality~\cite{Bell-RMP,Bell64}, the observation that no locally-causal theories~\cite{Bell04} can explain all the correlations between measurement outcomes obtained from certain entangled particles. Today, both phenomena are recognized as indispensable resources for various quantum information processing tasks, from computation~\cite{VidalPRL03,Jozsa:2003wj} to communication~\cite{Ekert91,Acin07,Vazirani14}, to name a few.

Quantum entanglement present in a {\em pure} state is {\em monogamous}~\cite{Ent-RMP,Coffman2000PRA}, i.e., it is impossible for composite systems, say, AB, to be in a pure entangled state while A, B, or AB together are also entangled with a third system C. Even though mixed-state entanglement can be shared, there is still a limitation on its shareability~\cite{Doherty:2014aa}, and some monogamy relations hold~\cite{Coffman2000PRA}. A very similar situation occurs for correlations between measurement outcomes observed in a Bell test: extremal Bell-nonlocal~\cite{Bell-RMP}, nonsignaling (NS)~\cite{PopescuFP94,BarrettPRA05} correlations must be monogamous~\cite{BarrettPRA05,MasanesPRA2006} but may otherwise be shareable~\cite{Collins04}. Even then, various tradeoffs on the amount of Bell violation are known (see, e.g.,~\cite{SG01,TV2006,Toner:2008aa,PB09,KPR+11,RH14,YLZ+24,CMR24} and references therein). 

When entanglement is not {\em strictly} monogamous, it can also exhibit a contrasting behavior. For example, there exist mixed bipartite quantum states for AB and BC, both entangled, such that all tripartite states for ABC compatible with these marginals must also return an entangled AC marginal state---a phenomenon called entanglement transitivity~\cite{Tabia2022}. For correlations in a Bell test, there is also a known example of NS correlation exhibiting the analogous nonlocality transitivity~\cite{Coretti2011}. However, as we show in this work, this example does not admit a quantum realization. In fact, before this work, it had remained unknown whether nonlocality transitivity could occur in {\em any form} within quantum theory.

Note that the existence of {\em nonlocality transitivity for correlations}~\cite{Coretti2011} can be used to argue against the plausibility of finite-speed causal influence models ~\cite{Scarani:2002aa,scarani_gisin_2005} for Bell-nonlocality. Even though such models accounting for the {\em quantum} violation of Bell inequalities have since been argued against theoretically using an alternative approach in~\cite{Bancal2012,Barnea2013}, it remains of interest to determine if nonlocality transitivity can occur in the quantum world. In particular, if proven impossible, the absence of such a feature in quantum theory makes it qualitatively different from {\em non}-quantum NS theories. In that case, physical theories allowing for nonlocality transitivity will be foil theories~\cite{Foil2016}.

Effectively, the search for a quantum example exhibiting nonlocality transitivity can be traced back to the work of~\cite{scarani_gisin_2005}, where they provided a tripartite quantum state having a Bell-nonlocal bipartite marginal. Tripartite quantum states having both~\cite{Collins04}, or even all Bell-nonlocal marginals~\cite{BV2012} are also known to exist. However, as we illustrate in~\cref{App:Candidates}, {\em none} of these are readily sufficient to illustrate nonlocality transitivity in quantum theory. In this work, we report a breakthrough in this long-standing problem by showing that the nonlocality of {\em quantum states} can indeed exhibit transitivity, thereby providing the first {\em quantum-realizable} notion of nonlocality transitivity.

We organize the rest of this paper as follows. \cref{Sec:Prelim} introduces the background for understanding Bell correlations, recalls from~\cite{Coretti2011} the notion of nonlocality transitivity of correlations, and explains its connection with finite-speed causal influence models~\cite{Scarani:2002aa,scarani_gisin_2005,Bancal2012}.  Then, in \cref{Sec:QNT}, we formalize notions of nonlocality transitivity in quantum theory at both the level of correlations and states. After that, by an example, we illustrate in \cref{Sec:FailNLTransitivity} the complication and subtleties involved in finding an example of nonlocality transitivity in quantum theory.  Our main results, consisting of analytic and numerical constructions of quantum states exhibiting nonlocality transitivity, are given in \cref{Sec:NLTransitivityStates}. Finally, we conclude in~\cref{Sec:Discussion} with further discussions of our results and possible directions for future research. All technical details, including a brief recapitulation of the Khot-Vishnoi nonlocal game~\cite{KhotVishnoi2005} and the relevant results from~\cite{Palazuelos2012,Cavalcanti2013}, are relegated to the Appendices.

\section{Preliminaries}\label{Sec:Prelim}

\subsection{Correlations in a Bell scenario}

Consider a bipartite Bell scenario where the measurement settings and outcomes of A and B are, respectively, labeled by $x$, $y$, and $a$, $b$. Furthermore, let us denote by $\vecPAB\coloneqq\{P(a,b|x,y)\}$ the collection of joint conditional probabilities observed in this bipartite Bell test. We say that the correlation $\vecPAB$ is nonsignaling~\cite{PopescuFP94,BarrettPRA05} (NS) if it satisfies:
\begin{subequations}\label{Eq:NS2}
\begin{align}
	\sum_{a} \PAB(a,b|x,y) = \sum_a \PAB(a,b|x',y),\label{Eq:PB}\\
	\sum_{b} \PAB(a,b|x,y) = \sum_b \PAB(a,b|x,y'),\label{Eq:PA}
\end{align}
\end{subequations}
for all $x,x',y,y'$ and all outcome random variables $a,b$ not summed over. When these conditions hold, we may define the marginal conditional probabilities arising from either side of \cref{Eq:PB,Eq:PA}, respectively, as $\vecPB$ and $\vecPA$. 
We denote the set of correlations respecting the NS conditions as $\NS$.

Correlations arising from local measurements on a shared quantum state are manifestly NS. To this end, we remind that a bipartite correlation $\vecPAB$ is said to be {\em quantum realizable} (within the tensor-product framework) if there exists a bipartite quantum state $\rho$ and local positive-operator-valued measures~\cite{QCI-book} (POVMs) $\{M^\tA_{a|x}\}$ and $\{M^\tB_{b|y}\}$ such that
\begin{equation}\label{Eq:P:Q}
	\PAB(a,b|x,y) = \tr(\rho\, M^\tA_{a|x}\otimes M^\tB_{b|y})\,\,\forall\,\,a,b,x,y.
\end{equation}
Hereafter, we refer to the set of quantum realizable correlations as $\Q$.

A celebrated discovery of Bell~\cite{Bell64} is that local-hidden-variable models cannot reproduce {\em all} $\vecPAB$ in the form of~\cref{Eq:P:Q}. In a bipartite Bell scenario, such models require that:
\begin{equation}\label{Eq:P:L}
	\PAB(a,b|x,y) = \sum_\lambda P_\lambda P(a|x,\lambda)P(b|y,\lambda)\,\,\forall\,\,a,b,x,y,
\end{equation}
for some normalized probability distributions $P_\lambda$ over the hidden variable $\lambda$ and local response functions $P(a|x,\lambda)$, $P(b|y,\lambda)\in\{0,1\}$. When a given $\vecPAB$ cannot be written in the form of \cref{Eq:P:L}, we say that it is Bell-nonlocal~\cite{Bell-RMP}, or simply nonlocal, and express this mathematically as $\vecPAB\not\in\cL$. A convenient way for manifesting this fact is that the given $\vecPAB$ violates a Bell inequality specified by $\vec{\beta}=\{\beta^{x,y}_{a,b} \}$:
\begin{equation}\label{Eq:BellIneq}
    \mathcal{I}:=\vec{\beta}\cdot\vecPAB=\sum_{x,y,a,b} \beta^{x,y}_{a,b} \PAB(a,b|x,y) \overset{\cL}{\le} B_{\vec{\beta}},
\end{equation}
where $\cL$ is the set of bipartite correlations that can be cast in the form of \cref{Eq:P:L}, 
and 
\begin{align}
	B_{\vec{\beta}}\coloneqq\max_{\vecPAB'\in\cL}\sum_{x,y,a,b} \beta^{x,y}_{a,b} \PAB'(a,b|x,y)
\end{align}
is the local upper bound associated to $\vec{\beta}$.

For the benefit of subsequent discussions, it is worth noting that the winning probability of a two-player nonlocal game~\cite{Cleve:IEEE:2004} can also be expressed as a linear combination of $\PAB(a,b|x,y)$ with $\beta^{x,y}_{a,b}\ge 0$. Then the local bound $B_{\vec{\beta}}$ is simply the best classical winning probability, usually denoted by $\omega_c$.

As with the case of entanglement, the phenomenon of nonlocality becomes considerably richer beyond the bipartite scenario. To this end, consider now a tripartite Bell scenario that also involves a third party C, whose measurement setting and outcome are, respectively, labeled by $z$ and $c$. By analogy, we denote by $\vecPABC\coloneqq\{P(a,b,c|x,y,z)\}$ the collection of joint conditional probabilities observed in this tripartite Bell test. We say that the correlation $\vecPABC$ is NS if it satisfies:
\begin{subequations}\label{Eq:NS3}
\begin{align}
	\sum_{a} \PABC(a,b,c|x,y,z) = \sum_a \PABC(a,b,c|x',y,z),\label{Eq:PBC}\\
	\sum_{b} \PABC(a,b,c|x,y,z) = \sum_b \PABC(a,b,c|x,y',z),\label{Eq:PAC}\\
	\sum_{c} \PABC(a,b,c|x,y,z) = \sum_c \PABC(a,b,c|x,y,z'),\label{Eq:PAB}
\end{align}
\end{subequations}
for all $x,x',y,y',z,z'$ and all outcome random variables $a,b,c$ not summed over. When \cref{Eq:NS3} holds, we may similarly define the marginal conditional probabilities arising from either side of \cref{Eq:PBC,Eq:PAC,Eq:PAB}, respectively, as $\vecPBC$, $\vecPAC$, and $\vecPAB$. To ease notation, we again denote the set of correlations respecting the NS conditions of \cref{Eq:NS3} as $\NS$, even though it is clear that the NS set to which $\vecPABC$ belongs is different from the one to which a bipartite NS correlation, such as $\vecPAB$, belongs (these vectors reside in real vector spaces of different dimensions). Similar remarks apply to all the other sets introduced above.

\subsection{Nonlocality transitivity of correlations and the untenability of finite-speed causal influence models}\label{Sec:NLTransitivity}

For any given NS $\vecPABC$, its bipartite marginals $\vecPAB$ and $\vecPBC$ are {\em uniquely} determined via \cref{Eq:NS3}. However, not all pairs of marginals $\vecPAB$ and $\vec{Q}_\text{BC}$ are {\em compatible}, i.e., even if they are {\em locally compatible} in the sense of both giving the same marginal at B ($\vecPB=\vec{Q}_\tB$), there is generally no guarantee that there exists a tripartite NS correlation $\vec{R}_\text{ABC}$ such that its marginals satisfy $\vec{R}_\text{AB}=\vecPAB$ and $\vec{R}_\text{BC}=\vec{Q}_\text{BC}$. For example, if $\vecPAB=\vec{Q}_\text{BC}$ is the {\em extremal} quantum correlation maximally violating the Clauser-Horne-Shimony-Holt (CHSH)~\cite{CHSH} Bell inequality, then it is known~\cite{BarrettPRA05,MasanesPRA2006,TV2006,Toner:2008aa} that there cannot exist any NS $\vec{R}_\text{ABC}$ giving both of these as its marginals.

Throughout, we {\em always} assume that the pair of marginals $\vecPAB$ and $\vecPBC$ are compatible, i.e., they can both be obtained from the same $\vecPABC\in\NS$ via \cref{Eq:NS3}. However, if we start with these marginals, there may also be other tripartite NS correlations $\vecPABC'\neq\vecPABC$ that return them via \cref{Eq:NS3}. 

We may now recall from~\cite{Coretti2011} the following definition.
\begin{definition}[Nonlocality transitivity of correlations~\cite{Coretti2011}]\label{Dfn:NTC}
{\em
    The pair of compatible marginals $\vecPAB$ and $\vecPBC$ exhibit {\em nonlocality transitivity} if 
    \begin{enumerate}
    \item They are both Bell-nonlocal, i.e., $\not\in\cL$. 
    \item For {\em all} tripartite NS correlations $\vecPABC'$ that return $\vecPAB$ and $\vecPBC$ as marginals via~\cref{Eq:NS3}, the corresponding marginal $\vecPAC'$ is also Bell-nonlocal.
    \end{enumerate}
}
\end{definition}
Importantly, \cref{Dfn:NTC} {\em does not} require any of the correlations $\vecPAB$, $\vecPBC$ (or $\vecPABC'$) to be quantum realizable via \cref{Eq:P:Q} (or its tripartite analog). 
More formally, if we denote by $\vecPAC'$ the AC marginal of $\vecPABC'$, then the pair of compatible marginal $\vecPAB$ and $\vecPBC$ satisfying \cref{Dfn:NTC} are such that
\begin{equation}\label{Eq:NTC}
	(\vecPAB, \vecPBC \not\in \cL) \land	(\vecPAC' \not\in \cL\,\,\forall\,\,\vecPABC'\in\CsABBC{\NS}),
\end{equation}
where $\land$ is a shorthand for the logical AND, and $\CsABBC{\NS}$ is the collection of all tripartite correlations $\vecPABC'$ in $\NS$ returning $\vecPAB$ and $\vecPBC$ as marginals via \cref{Eq:NS3}.

\begin{figure}[ht]
    \centering    \includegraphics[width=0.85\columnwidth]{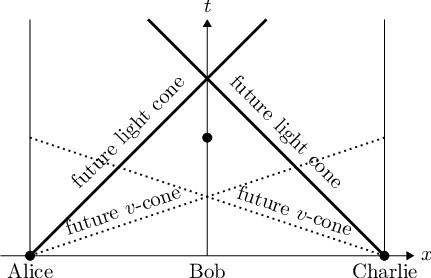}
\caption{Spacetime diagram illustrating a hypothetical tripartite Bell experiment where the measurement events of Alice, Bob, and Charlie occur, respectively, at the left, middle, and right spacetime point (represented by $\bullet$). The solid line extending from Alice's (Charlie's)  measurement event marks the boundary of the future light cone of this event; likewise,  the dotted lines extending from these measurement events mark the boundary of their future $v$-cone (i.e., the causal cone defined by the hypothetical finite-speed causal influence). In this reference frame, since Alice and Charlie measure simultaneously, $\vecPAC$ must be Bell-local according to the working hypothesis of the finite-speed $v$-causal model. On the contrary, Bob's measurement event lies inside the future $v$-cone of the other measurement events, and hence can be {\em Bell-nonlocal}.  }
\label{Fig:spacetime}
\end{figure}

In~\cite{Coretti2011}, \cref{Dfn:NTC} was introduced to argue against finite-speed causal influence models~\cite{Scarani:2002aa,scarani_gisin_2005} ({\em a.k.a.} $v$-causal models~\cite{Bancal2012,Barnea2013}) for explaining Bell-nonlocal correlations. These models assume that (in analogy to hidden variables) finite-speed, superluminal hidden causal influences are the mechanisms by which nonlocal correlations are established in nature. Then, by providing an example satisfying \cref{Dfn:NTC}, one can show that the three assumptions {\em cannot} hold simultaneously (see also~\cite{scarani_gisin_2005}):
\begin{enumerate}
\item $v$-causal model generates Bell-nonlocal correlations only if the finite-speed causal influence arrives in time.
\item $v$-causal model only produces NS correlations.
\item $v$-causal model generates Bell-nonlocal $\vecPAB$ and Bell-nonlocal $\vecPBC$ exhibiting nonlocal transitivity, cf.~\cref{Dfn:NTC}, in the spacetime configuration shown in~\cref{Fig:spacetime}.
\end{enumerate}
The first assumption above is the working hypothesis of $v$-causal models for finite $v$; the second assumption embodies the requirement that these influences are also hidden in the sense that they do not allow for superluminal communication. A priori, these two assumptions need not be incompatible. However, when we apply assumption 1 to the spacetime configuration of \cref{Fig:spacetime}, the model requires that $\vecPAC$ is Bell-local, which contradicts assumption 3. In other words, the existence of $\vecPAB,\vecPBC$ satisfying \cref{Dfn:NTC} (if observed) implies undesirable features of such models: either such causal influences must propagate at {\em infinite} speed, or they must violate assumption 2, and hence, allow for superluminal communication even when one does not have access to them.

In~\cite{Coretti2011}, the authors provided an example of NS marginals $\vecPAB$ and $\vecPBC$ (involving quaternary values for $x,y$, and $z$ but binary values for $a$ and $b$) such that for all NS $\vecPABC'$ returning these bipartite marginals, the corresponding $\vecPAC'$ must also be Bell-nonlocal. However, they have left  in~\cite{Coretti2011} the quantum realizability of their example as an open problem. In~\cref{App:Postquantum}, we provide evidence showing that their example is indeed {\em not} quantum realizable, i.e., $\vecPAB,\vecPBC\not\in\Q$, thereby answering their open problem to the {\em negative}. Hence, unless their {\em non}-quantum correlations can somehow be realized in our world, which seems implausible given the work of~\cite{BBL+06,Pawowski:2009aa,Navascues:2009aa,SGA+18} (see also references in~\cite{Bell-RMP}), the physical implications of their example are unclear. In particular, it naturally raises the question of whether nonlocality transitivity is a phenomenon that can occur within quantum theory or whether it is a feature compatible only with theories that have to embrace {\em non}-quantum correlations.

\section{Towards quantum nonlocality transitivity}
\label{Sec:QNT}

Given this state of affairs, it should be clear that, even from a foundational perspective alone, it remains of strong interest to determine if nonlocality transitivity can occur in some form within quantum theory. To this end, we introduce below our formulations of a quantum example and explain why they are appropriate in the present context.

\subsection{Transitivity at the level of quantum correlations}

Evidently, a quantum example must involve input marginals $\vecPAB$ and $\vecPBC$ that are not only NS but are also quantum realizable via \cref{Eq:P:Q}. 
By imposing on \cref{Dfn:NTC} the requirement that these are marginals derived from a $\vecPABC\in\Q$ thus leads to the following definition.\footnote{Instead of \cref{Dfn:NTCQ}, one may be tempted to define a quantum example as one that satisfies $\vecPAB,\vecPBC\in\Q$ in addition to \cref{Dfn:NTC}. However, as mentioned above, requiring $\vecPAB$ and $\vecPBC$ to be quantum realizable generally does not guarantee that they are (quantum) compatible. See~\cref{App:OtherDefinitions} for further discussions.}
\begin{definition}[Nonlocality transitivity of quantum correlations]\label{Dfn:NTCQ}
	{\em
    The pair of compatible marginals $\vecPAB$ and $\vecPBC$ exhibits {\em quantum nonlocality transitivity} if 
    \begin{enumerate}
    \item They satisfy \cref{Dfn:NTC}.
    \item They can be recovered as the marginals of a tripartite quantum correlation $\vecPABC$.
    \end{enumerate}
    }
\end{definition}
Note that the second condition above implies that both $\vecPAB$ and $\vecPBC$ are quantum realizable via \cref{Eq:P:Q}.
Hence, the pair of compatible $\vecPAB$ and $\vecPBC$ satisfy \cref{Dfn:NTCQ} if and only if
\begin{equation}\label{Eq:NTCQ}
	(\vecPAB, \vecPBC \not\in \cL) \land (\vecPAC' \not\in \cL\,\,\forall\,\,\vecPABC'\in\CsABBC{\NS}) \land (\CsABBC{\Q}\neq\emptyset),
\end{equation}
where $\CsABBC{\Q}$ is the collection of all tripartite correlations $\vecPABC'$ in $\Q$ returning $\vecPAB$ and $\vecPBC$ as marginals via \cref{Eq:NS3} whilst $\emptyset$ denotes the empty set.
Since $\Q\subset\NS$, any $\vecPAB$, $\vecPBC$ satisfying \cref{Eq:NTCQ} must also satisfy the weaker set of conditions
\begin{equation}\label{Eq:NTCQ0}
	(\vecPAB, \vecPBC \not\in \cL)  \land (\vecPAC' \not\in \cL\,\,\forall\,\,\vecPABC'\in\CsABBC{\Q}).
\end{equation}

It may be tempting to take \cref{Eq:NTCQ0} as the defining conditions for a weaker notion of nonlocality transitivity. However, establishing an example satisfying \cref{Eq:NTCQ0} can only lead to a weaker conclusion. Specifically, even if we insist that the causal influence propagates at a finite speed, the existence of marginals $\vecPAB$, $\vecPBC$ satisfying \cref{Eq:NTCQ0} will only imply that the tripartite distribution $\vecPABC$ produced by these models is necessarily {\em not} quantum, rather than {\em signaling}. Since the $v$-causal model---in contrast with quantum theory---produces correlations that depend on the spacetime configuration, the deviation of its prediction from quantum theory is arguably not surprising after all.   

\subsection{Transitivity at the level of quantum states}

The search for a quantum example fulfilling~\cref{Dfn:NTCQ} is far from trivial as it requires one to work with nonconvex sets (the complement of $\cL$ among all quantum correlations for AB, BC, and AC). More precisely, for a quantum example that fulfills \cref{Dfn:NTCQ} to exist, we must have $\vecPABC\in\Q$ such that all its bipartite marginals $\vecPAB$, $\vecPBC$, and $\vecPAC$ are Bell-nonlocal. For convenience, we say that such a $\vecPABC$ exhibits {\em quantum nonlocality for all marginal correlations} (QNAMC). While we are unaware of any existing example of $\vecPABC$ exhibiting QNAMC, their construction---if there is no restriction on the number of measurement settings and outcomes---turns out to be straightforward, as we shall demonstrate in~\cref{Sec:FailNLTransitivity}. Moreover, 
we show below that another necessary condition for such examples to exist can also be fulfilled. 

To this end, let us follow the usual convention and refer to a quantum state $\rho$ as {\em nonlocal} if it gives nonlocal correlations via a judicious choice of POVMs via \cref{Eq:P:Q}. Moreover, as with the case of correlations, we say that two bipartite states $\rAB$ and $\rBC$ are {\em compatible} if there exists at least one tripartite quantum state $\rho_\text{ABC}$ that returns $\rAB$ and $\rBC$ as reduced states (via partial tracing). Then, the following Lemma can be shown to hold.
\begin{lemma}\label{Lem:CortoState}
	For a quantum example that fulfills \cref{Dfn:NTCQ} [and hence \cref{Eq:NTCQ0}] to exist, the corresponding pair of density matrices $\rAB$ and $\rBC$ giving rise to the $\vecPAB$ and $\vecPBC$ must satisfy the transitivity property specified in~\cref{Dfn:NLTRho}.
\end{lemma}
\begin{definition}[Nonlocality transitivity of quantum states]\label{Dfn:NLTRho}
	{\em 
    The pair of compatible marginal states $\rAB$ and $\rBC$ exhibit {\em nonlocality transitivity} if 
    \begin{enumerate}
    \item They are both nonlocal. 
    \item For all tripartite state $\rho'_\text{ABC}$ that returns $\rAB$ and $\rBC$ as reduced states, the corresponding reduced state $\rho'_\text{AC}$ is also nonlocal.
    \end{enumerate}
    }
\end{definition}

Let $\mathfrak{C}$ be the set of all states in ABC that return both $\rAB$ and $\rBC$ by partial tracing, respectively, over C and A. Moreover, let $\D_\cL$ be the set of {\em local} (i.e., non-Bell-inequality-violating) states. Then the conditions of \cref{Dfn:NLTRho}, central to our discussion, can be expressed as follows for a pair of compatible states $\rAB$ and $\rBC$:
\begin{equation}\label{Eq:NLTransitivityRho}
	(\rAB, \rBC \not\in \D_\cL) \land (\tr_\tB\rABC' \not\in \D_\cL\,\,\forall\,\,\rABC'\in\mathfrak{C}).
\end{equation}

\begin{proof}
To prove Lemma~\ref{Lem:CortoState}, let $\vecPABC\in\Q$ be a tripartite quantum correlation whose marginals $\vecPAB$ and $\vecPBC$ satisfy the requirement of \cref{Eq:NTCQ0}. Since $\vecPABC$ is quantum realizable, there exists a density operator $\rABC$ and local POVMs $\{M^\tA_{a|x}\},\{M^\tB_{b|y}\},\{M^\tC_{c|z}\}$ such that
\begin{equation}\label{Eq:Q3}
	\PABC(a,b,c|x,y,z) = \tr(\rABC\,M^\tA_{a|x}\otimes M^\tB_{b|y}\otimes M^\tC_{c|z})
\end{equation}
for all $a,b,c,x,y,z$. From~\cref{Eq:NS3,Eq:Q3}, and the normalization condition $\sum_c M^\tC_{c|z}=\id_\tC$, we get
\begin{equation}\label{Eq:BipartiteMarginals}
\begin{split}
	\PAB(a,b|x,y) &= \sum_c \tr(\rABC\,M^\tA_{a|x}\otimes M^\tB_{b|y}\otimes M^\tC_{c|z}),\\
	 &= \tr(\rAB\,M^\tA_{a|x}\otimes M^\tB_{b|y}),
\end{split}
\end{equation}
and similar expressions for $\vecPBC$ and $\vecPAC$ in terms, respectively, of $\rBC$ and $\rAC$. From the requirements that $\vecPAB,\vecPBC\not\in\cL$ and the definition of $\D_\cL$, we deduce immediately that $\rAB,\rBC\not\in\D_\cL$. Now, let us suppose there exists another density operator $\rABC'$ such that its reduced states satisfy $\rAB'=\rAB$, $\rBC'=\rBC$, but $\rAC'\in\D_\cL$, i.e., $\rAB, \rBC$ do not satisfy the requirements of \cref{Dfn:NLTRho}. Then, by adopting exactly the same local measurements, we get $\vecPAB'=\vecPAB$, $\vecPBC'=\vecPBC$, but
\begin{equation}\label{Eq:DeducedMarginals}
\begin{split}
	\PAC'(a,c|x,z) &= \sum_b \tr(\rABC'\,M^\tA_{a|x}\otimes M^\tB_{b|y}\otimes M^\tC_{c|z}),\\
	 &= \tr(\rAC'\,M^\tA_{a|x}\otimes M^\tC_{c|z}).
\end{split}
\end{equation}
However, since $\rAC'\in\D_\cL$, we must have $\vecPAC'\in\cL$, which contradicts the second requirement of \cref{Dfn:NTC}, and hence also the first requirement of \cref{Dfn:NTCQ}. Therefore, as claimed, any example complying with \cref{Dfn:NTCQ} must also comply with \cref{Dfn:NLTRho}.
\end{proof}

\begin{figure}[t!hbp]
    \centering    \includegraphics[width=0.9\columnwidth]{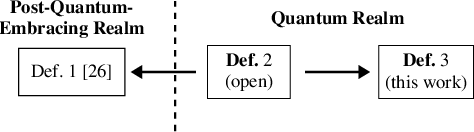}
\caption{Summary of the relationships between the various Definitions (Def.) discussed. A solid vector pointing from A to B indicates that if an example meets the requirements of A, there must also be an example meeting those of B. In other words, an example for B must exist for that of A to exist. Notably, since $\Q\subsetneq\NS$ while~\cref{Dfn:NTC} only involves the NS constraints of \cref{Eq:NS3}, examples satisfying~\cref{Dfn:NTC} may be {\em post-quantum} (i.e., even more strongly correlated than that allowed in quantum theory) and hence
do not necessarily meet the criteria for~\cref{Dfn:NTCQ}. Indeed, as we show in~\cref{App:Postquantum}, the marginals of the known example~\cite{Coretti2011} are {\em non}-quantum, failing to satisfy~\cref{Dfn:NTCQ} for the nonlocality transitivity in the quantum realm.
In this work, we provide the first examples for \cref{Dfn:NLTRho}.}
\label{Fig:RelateDefinitions}
\end{figure}

We summarize the relations between all preceding Definitions in \cref{Fig:RelateDefinitions}. At this point, it is worth reminding that entanglement is  {\em necessary but generally insufficient}~\cite{Werner:PRA:1989} for exhibiting Bell-nonlocality. However, determining if a given entangled quantum state can violate any Bell inequality is generally a high-dimensional variational problem to which only a partial solution is known, see, e.g.,~\cite{Terhal03,LD07,HirschPRL16,CavalcantiPRL16,Hsieh2016} and references therein. In fact, even if we restrict our attention to two-qubit states, little is known beyond the pioneering result by Horodecki {\em et al.}~\cite{HHH95}. Hence, even though there are known examples of tripartite quantum states giving entangled marginals (e.g., the three-qubit state $W$ state~\cite{Dur2000}), they do not necessarily translate to results on nonlocality transitivity.
 
\begin{figure}[h!]
    \centering    \includegraphics[width=0.5\columnwidth]{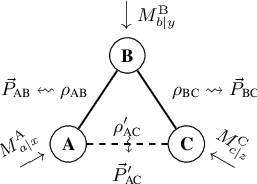}
    \caption{For the problem of nonlocality transitivity in a tripartite setting, we consider the AB and BC marginals to be given (solid lines) and compatible, whilst the nonlocality property of the AC marginal (dashed line) is to be inferred. For marginal density matrices $\rAB$ and $\rBC$ of $\rABC$ to exhibit the nonlocality transitivity of \cref{Dfn:NLTRho}, all $\rABC'$ consistent with these marginals must give nonlocal $\rAC':=\tr_\tB\rABC'$. Hence, {\em all} bipartite marginals of $\rABC'$ are nonlocal, i.e., $\rABC'$ {\em must} lie outside the union of the three biseparable sets with {\em fixed} bipartitions A|BC, B|AC, and C|AB. However, lying outside the convex hull of these three sets, i.e., being genuinely tripartite entangled, is generally insufficient to fulfill the requirement of \cref{Dfn:NLTRho} either. By performing local measurements on A, B, and C, with POVMs given, respectively, by $\{M^\tA_{a|x}\}$, $\{M^\tB_{b|y}\}$, and $\{M^\tC_{c|z}\}$, we obtain a tripartite quantum correlation $\vecPABC$ with marginals $\vecPAB$ and $\vecPBC$. In general, these marginals may also be compatible with other $\vecPAC'$ (derived from $\rAC'$ or other NS correlation). If $\vecPAB$, $\vecPBC$, and {\em all} $\vecPAC'$ derived from NS $\vecPABC'$ are Bell-nonlocal, then the marginal correlations also serve as an example satisfying~\cref{Dfn:NTCQ}.}
    \label{Fig1:Marginals}
\end{figure}

Note further that any tripartite state $\rABC$ whose marginals $\rAB$ and $\rBC$ satisfy \cref{Dfn:NLTRho} must be such that all its bipartite marginals are Bell-nonlocal. However, before this work, the only known example~\cite{BV2012} exhibiting this phenomenon of {\em quantum nonlocality for all marginal states} (QNAMS) does not exhibit nonlocality transitivity (see~\cref{App:BVExample} for details). With some thoughts, one also realizes~\cite{PrivateMarwan} that any $\rABC$ fulfilling \cref{Dfn:NLTRho} {\em cannot} be separable with respect to any of the bipartitions; otherwise, at least one of the marginals will be separable, and hence~\cite{Werner:PRA:1989} local. However, taking the three-qubit Greenberger-Horne-Zeilinger state as an example, it is evident that demanding $\rABC$ to be genuinely tripartite entangled also does not help to ensure that \cref{Dfn:NLTRho} holds for its marginals (see also~\cite{Toth2007,WBA+12,LCBG14,Navascues2021} and references therein). These observations are summarized in \cref{Fig1:Marginals}.

\section{An intuitive but unsuccessful attempt via the construction of QNAMS and QNAMC}
\label{Sec:FailNLTransitivity}

\begin{figure}[h!]
    \centering    \includegraphics[width=0.35\columnwidth]{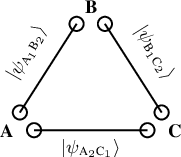}
\caption{Schematic showing an intuitive but unsuccessful attempt to produce a quantum example of nonlocality transitivity.}
\label{Fig:NaiveAttempt}
\end{figure}

Despite all the remarks above, one may still feel that it is trivial to construct examples of quantum nonlocality transitivity by putting together known examples of quantum states violating a Bell inequality. Next, we demonstrate by an example why the problem is more subtle than it may seem, even though one can indeed get very close to establishing an example satisfying both~\cref{Dfn:NTCQ} and \cref{Dfn:NLTRho}. To this end, let $i,j\in \{\tA, {\rm B}, {\rm C}\}$, $k,\ell\in\{1,2\}$, with $i\neq j$ and $k\neq\ell$, and imagine that A, B, and C each share a Bell state (written on a rotated basis)~\cite{BNS+15}
\begin{equation}
    \ket{\psi_{i_kj_\ell}} = \frac{1}{\sqrt{2}} \left[\cos\frac{\pi}{8}(\ket{00}-\ket{11})+\sin\frac{\pi}{8}(\ket{01}+\ket{10} \right],
\end{equation}
with each of the other two parties, as illustrated in~\cref{Fig:NaiveAttempt}.

It is easy to verify that if any of the two parties, say, A and B, locally measure $\sigma_z$ ($\sigma_x$) for their zeroth (first) measurement on their shared entangled state $\ket{\psi_\text{A$_1$B$_2$}}$, they obtain a correlation $\vecP_{\tA_1\tB_2}$ that violates the CHSH-Bell inequality~\cite{CHSH} maximally, i.e, 
\begin{equation}
	\sum_{\substack{x_1,y_2, \\ a_1,b_2=0,1}}
	\!\!\!(-1)^{a_1+b_2+x_1y_2} P_{\tA_1\tB_2}(a_1,b_2|x_1,y_2) = 2\sqrt{2} \not\le 2, 
\end{equation}	
where we have identified the $+1$ ($-1$)-outcome of the dichotomic measurement with the output bit $0$ ($1$). In fact, if we measure all particles in exactly the same manner, the bipartite marginal, say, $\vecPAB$, takes the form of
\begin{equation}
	\PAB(a_1,a_2,b_1,b_2|x_1,x_2,y_1,y_2) 
	= \frac{1}{16}\left[ 1+\frac{(-1)^{a_1+ b_2+x_1y_2}}{\sqrt{2}}\right].
\end{equation}	
Consequently, $\vecPAB$ violates the four-setting, four-outcome ``extension'' of the CHSH-Bell inequality $\mathcal{S}\overset{\cL}{\le} 8$, where
\begin{align}
	\mathcal{S}:&=\!\!\!\!\!\!\sum_{\substack{x_1,x_2,y_1,y_2, \\ a_1,a_2,b_1,b_2=0,1}}
	\!\!\!\!\!\!\!\!\!\!\!(-1)^{a_1+b_2+x_1y_2} \PAB(a_1,a_2,b_1,b_2|x_1,x_2,y_1,y_2),\nonumber\\
	&=4\sum_{\substack{x_1,y_2, \\ a_1,b_2=0,1}}
	\!\!\!(-1)^{a_1+b_2+x_1y_2} P_{\tA_1\tB_2}(a_1,b_2|x_1,y_2). 
\end{align}	
By symmetry, we see that $\vecPBC$ and $\vecPCA$ also violate the same Bell inequality. Hence, the described quantum strategy gives an example of $\vecPABC$ exhibiting QNAMC, which we remind is a pre-requisite for establishing an example for~\cref{Dfn:NTCQ}.

Even then, we shall see that it does not constitute a quantum example of nonlocality transitivity because the quantum state involved (see~\cref{Fig:NaiveAttempt})
\begin{equation}\label{Eq:OriginalRho}
	\rABC=\proj{\psi_\text{A$_1$B$_2$}}\!\otimes\!\proj{\psi_\text{B$_1$C$_2$}}\!\otimes\!\proj{\psi_\text{A$_2$C$_1$}}
\end{equation} 
fails to fulfil the requirement of \cref{Dfn:NLTRho}. More explicitly, for each pair $i=\tA,j=\tB$, or $i=\tB,j=\tC$, or $i=\tC,j=\tA$, we see that the shared density matrix is 
\begin{equation}\label{Eq:MarginalRho}
	\rho_{ij} = \proj{\psi_{i_1j_2}}\otimes\frac{\id_{i_2}}{2}\otimes\frac{\id_{j_1}}{2},
\end{equation}
where we have used the fact that the reduced density matrix of a Bell state is the maximally mixed qubit state $\frac{\id}{2}$. Note that a {\em different} global state $\vrABC\neq\rABC$ compatible with the reduced states $\rho_\text{AB}$ and $\rho_\text{BC}$ of \cref{Eq:MarginalRho} is 
\begin{equation}\label{Eq:AlternativeRho}
	\vrABC = \proj{\psi_\text{A$_1$B$_2$}}\otimes\proj{\psi_\text{B$_1$C$_2$}}\otimes\frac{\id_{\tA_2}}{2}\otimes\frac{\id_{\tC_1}}{2},
\end{equation}
which gives the separable, hence Bell-local, AC reduced state
\begin{equation}\label{Eq:DeducedMarginalRho}
	\vrAC = \frac{\id_{\tA_1}}{2}\otimes\frac{\id_{\tA_2}}{2}\otimes\frac{\id_{\tC_1}}{2}\otimes\frac{\id_{\tC_2}}{2}.
\end{equation}
In other words, even though the $\rABC$ of \cref{Eq:OriginalRho} leads to Bell-nonlocal $\vecPAB, \vecPBC$, and $\vecPAC$, the fact that its marginals $(\rAB, \rBC)$ are compatible with a separable $\vrAC$, \cref{Eq:DeducedMarginalRho}, means that they do not satisfy \cref{Dfn:NLTRho}. Hence, by Lemma~\ref{Lem:CortoState}, $(\vecPAB, \vecPBC)$ will never be a quantum example satisfying \cref{Dfn:NTCQ}.

Although the above construction fails to give a quantum state with reduced states showing nonlocality transitivity, one may have noticed that the tripartite state of \cref{Eq:OriginalRho} does exhibit QNAMS, which is a {\em prerequisite} for showing nonlocality transitivity. Indeed, with some thoughts, one notices a straightforward construction of an $n$-partite quantum state for $\tA_1\tA_2\cdots\tA_n$ that exhibits QNAMS for arbitrary $n>2$. Specifically, for each $k=2,3,\cdots, n-1$, take a $k$-partite nonlocal state $\rho_{i_1i_2\cdots i_k}$, with $i_1,i_2,\cdots,i_k\in\{1,2,\cdots,n\}$ being {\em non-repeating} indices of the $k$ parties, then the global state (with non-repeating indices $i_1,i_2,\cdots, i_k$)
\begin{equation}
	\bm{\rho}_{\tA_1\tA_2\cdots\tA_n} = \bigotimes_{k=2}^{n-1} \bigotimes_{i_1,i_2,\cdots,i_k} \rho_{i_1i_2\cdots i_k}
\end{equation}
is easily seen to exhibit QNAMS. Moreover, by allowing each subsystem to have its own measurement settings and outcomes, cf.~\cref{Fig:NaiveAttempt} and \cref{Fig1:Marginals} with $M^\tA_{a|x} = M^{\tA_1}_{a_1|x_x}\otimes M^{\tA_2}_{a_2|x_x}$, etc.,  we can analogously construct an $n$-partite correlation exhibiting QNAMC. After all, each $k$-particle state $\rho_{i_1i_2\cdots i_k}$ is, {\em by assumption}, nonlocal. Thus, via a judicious choice of local POVMs on each of these $k$-particle states, we must be able to manifest the nonlocality of the $k$-particle marginal correlation that results from $\rho_{i_1i_2\cdots i_k}$.

However, for the case of $n=3$, the above construction inherits the same weakness of \cref{Eq:OriginalRho}, i.e., one can always find a separable state $\vrAC$ compatible with the marginals $\bm{\rho}_\text{AB}$ and $\bm{\rho}_\text{BC}$, thus rendering it useless for demonstrating nonlocality transitivity.

\section{Nonlocality of Quantum States can be Transitive}
\label{Sec:NLTransitivityStates}

We now present two very different constructions of quantum states satisfying~\cref{Dfn:NLTRho}, thereby showing that nonlocality transitivity can occur in some form in quantum theory. By definition, for any marginal states of AB and BC exhibiting nonlocality transitivity, we must be able to infer the properties of the state of AC from those of AB and BC, cf.~\cref{Fig1:Marginals}.  In particular, if the latter uniquely determines the global compatible state, then the property of AC can also be deduced accordingly. Both constructions presented here exploit this uniqueness property of the input bipartite marginals.

\subsection{An analytic construction}

\subsubsection{Copies of two-body marginals of the $W$-state determine the global state uniquely}
\label{sec:uniqueness}

Let us begin by considering the $n$-qubit $W$ state~\cite{Dur2000}:
\begin{equation}
	\ket{W_n}=\frac{1}{\sqrt{n}}(\ket{10\cdots 0}+\ket{010\cdots 0}+\cdots\ket{0\cdots 01}).
\end{equation}
Any two-qubit reduced state of $\ket{W_n}$ is easily shown to be:
\begin{equation}\label{Reduced_state}
    \tau_n\coloneqq \frac{2}{n}\proj{\Psi^+}+\frac{n-2}{n}\proj{00},
\end{equation}
where $\ket{\Psi^\pm}=\frac{1}{\sqrt{2}}(\ket{01}\pm\ket{10})$. Conversely, for any tree graph~\cite{BenderWilliamson2010} with $n$ vertices such that any two vertices connected by an edge are described by $\tau_n$, it is known~\cite{Parashar.PRA.2009,Wu2014} that $\ket{W_n}$ is the unique compatible global state.

A generalization of this result involving a one-parameter family of reduced states can be found in~\cite{Tabia2022}. In the following, we present a different generalization of the above uniqueness result involving identical copies of $\tau_n$.
\begin{lemma}\label{Lem:UniqueGlobalState}
	For any tree graph with $n$ vertices such that any two vertices connected by an edge are described by $\tau_n^{\otimes k}$, the only global state compatible with these marginals is $k$ copies of the $n$-qubit $W$-state $\ket{W_n}^{\otimes k}$.
\end{lemma}

We give the proof of Lemma~\ref{Lem:UniqueGlobalState} in \cref{App:ProofUniqueness}. Note that the extension of uniqueness results from a single copy to multiple copies of the marginals of pure states has also been discussed in~\cite{SC23}. However, their result requires the specification of all $\binom{n}{2}$ two-body marginals, whereas ours only requires the specification of $(n-1)$ two-body marginals.

\subsubsection{Bell-nonlocality transitivity of quantum states}
\label{Sec:BNLTransitivity}

We are now ready to present our first examples of quantum nonlocality transitivity, which consist of examples of marginal states exhibiting nonlocality transitivity for quantum states, cf.~\cref{Dfn:NLTRho}.

\begin{theorem}[Bell-nonlocality transitivity]\label{Result:NonlocalityTransitivity}
	For every integer $k$ larger than some threshold value $k_c\in\mathbb{N}$, there exist compatible nonlocal $\rAB, \rBC$ such that for every $\rABC'$ acting on $[(\mathbb{C}^2)^{\otimes k}]\otimes[(\mathbb{C}^2)^{\otimes k}]\otimes [(\mathbb{C}^2)^{\otimes k}]$ that return $\rAB$ and $\rBC$ as marginals, the corresponding reduced state $\rAC'$ must be nonlocal.
\end{theorem}

To prove \cref{Result:NonlocalityTransitivity}, we take {\em multiple copies} of the two-qubit reduced state of $\ket{W_3}$, i.e., $\tau_3^{\otimes k}$ as the marginals $\rAB$ and $\rBC$. Then, Lemma~\ref{Lem:UniqueGlobalState} implies that the global state must be $\ket{W_3}^{\otimes k}$, and hence any compatible $\rAC'$ must be $\tau_3^{\otimes k}$. To demonstrate the nonlocality of these reduced states, we make use of the Khot-Vishnoi (KV) nonlocal game~\cite{KhotVishnoi2005} and the construction of Bell violation by tensoring first given by Palazuelos~\cite{Palazuelos2012} and further improved by Cavalcanti {\em et al.}~\cite{Cavalcanti2013}. 

\begin{figure}[h!]
    \centering    \includegraphics[width=0.5\columnwidth]{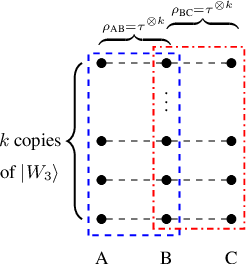}
\caption{In our analytic construction of the nonlocality transitivity of quantum states, we take $\rAB$ (systems enclosed within the blue dashed rectangle) and $\rBC$ (systems enclosed within the red dashed-dotted rectangle) to be $\tau_3^{\otimes k}$, i.e., $k$ copies of the reduced density matrix of the three-qubit $W$-state. By Lemma~\ref{Lem:UniqueGlobalState}, the only global state of ABC compatible with these marginals is $\ket{W_3}^{\otimes k}$, and thus the AC marginal {\em must be}  $\tau_3^{\otimes k}$. Since $\tau$ has a fully-entangled fraction~\cite{HorodeckiMP1999} given by $\frac{2}{3}$, all these marginals provably violate the KV Bell inequality for $k\ge 29$, thus showing the nonlocality of $\rAB$, $\rBC$, all compatible $\rAC'$, and hence the nonlocality transitivity of $\rAB$ and $\rBC$. See~\cref{App:ProofNLTransitivityRho} for details.}
\label{Fig:ProofIllustration}
\end{figure}

In particular, it follows from the result of~\cite{Cavalcanti2013} that $\ket{W_3}^{\otimes k}$ exhibits QNAMS for sufficiently large $k$, which, when combined with Lemma~\ref{Lem:UniqueGlobalState}, concludes the proof. For completeness, we briefly recapitulate in \cref{App:KV} the KV game and some details of the construction by Cavalcanti {\em et al.}~\cite{Cavalcanti2013}. In~\cref{Fig:ProofIllustration}, we sketch the main ingredients of the proof and provide its details in~\cref{App:ProofNLTransitivityRho}.

\subsection{Numerical construction from Haar-random pure states}
\label{Sec:NumExamples}

Admittedly, our analytic construction above is far too complicated for any experimental implementation or practical application. We have also not found any nonlocality transitivity example by numerically optimizing the Bell violation of several obvious candidate quantum states from the literature, see~\cref{App:Candidates} for details. However, the discovery of generic entanglement transitivity among Haar-random tripartite pure states---observed numerically in~\cite{Tabia2022}, proved very recently in~\cite{Liu2024}---has provided many more candidates to our disposal.

To this end, we have generated Haar-random three-qudit pure states $\ket{\psi_d}\in\mathbb{C}^d\otimes\mathbb{C}^d\otimes\mathbb{C}^d$ with local dimensions $d=2,3,4$, and $5$, computed their bipartite reduced states, and optimized over $M=2, 3$ binary-outcome POVMs for each party to search for marginals $\rAB$ and $\rBC$ satisfying \cref{Dfn:NLTRho}. No such example is found among all the generated three-qubit states $\ket{\psi_2}$. In fact, {\em none} of them seems to exhibit QNAMS. However, the situation changes drastically when we consider Haar-random three-qutrit pure states $\ket{\psi_3}$, where we find about $11.4\%$ of them exhibiting QNAMS. Together with the fact that these marginals uniquely determine the global tripartite state (pure or mixed),\footnote{This is anticipated from the uniqueness results of~\cite{CDJ+13} and can be verified numerically from a fidelity-minimization computation (see~\cite{Tabia2022}).} we thus conclude that these marginals indeed exhibit nonocality transitivity. In~\cite{githubCode}, we provide the explicit form of some of these examples.

Does this phenomenon become more ubiquitous if we increase the dimension $d$ even further? Our preliminary investigation, see~\cref{Tbl:NLTRho:Summary}, suggests this is not necessarily true. However, a more comprehensive study using, e.g., a more extensive set of Bell inequalities, will be required to answer this question affirmatively.

\begin{table}[h]
\centering
\renewcommand{\arraystretch}{1.3}
\begin{tabular}{c|c|c|c|ccc|c}
$d$ & $M$ & Samples & Guesses & None (\%) & One (\%) & Two (\%) & All (\%) \\ 
\midrule
2 & 2 & 100000  & 36 & 10.73 & 89.27 & 0 & 0 \\ 
2 & 3 & 100000 & 81 & 10.51 & 88.97 & 0.51  & 0\\ \hline
3 & 2 & 100000& 256  &  12.81 & 41.42 & 34.46  & 11.31\\ 
3 & 3 & 100000 & 576 & 12.74  & 41.37 & 34.48 & 11.41 \\ \hline
4 & 2 & 10000 & 900 & 73.52 & 22.95 & 3.30 & 0.23 \\
4 & 3 & 10000 & 2056 & 73.54 & 22.94 & 3.29 & 0.23 \\ \hline
5 & 2 & 1000  & 2304 & 99.8 & 0.2 & 0 & 0 \\
5 & 3 & 1000 & 5184 & 99.8 & 0.2 & 0 & 0 \\
\end{tabular}
\caption{Fraction of three-qudit Haar-random pure states whose bipartite marginals (reduced states) show nonlocality. From left to right, the first three columns show the local dimension $d$ of the randomly generated pure states, the number of measurement settings $M$ employed in checking for the nonlocality of the reduced states, the number of samples generated in these tests, and the number of initial guesses used in optimizing the Bell violation via the algorithm given in~\cite{LD07}. Further to the right, we show, respectively, the fraction of samples where none, one, two, and all of the bipartite marginals are found to be nonlocal.}
\label{Tbl:NLTRho:Summary}
\end{table}

\section{Discussion}\label{Sec:Discussion}

Quantum nonlocality has always been a fascinating topic in the studies of quantum foundations~\cite{Liang:PRep} and, more recently, device-independent quantum information~\cite{Bell-RMP}. In this work, we provide explicit examples showing that nonlocality can be transitive for quantum states, thereby providing the {\em first quantum-realizable} notion of nonlocality transitivity, affirmatively answering a problem that has remained open since 2011. While our {\em analytic} examples appear challenging in their realization, it is interesting to note that the reduced states of {\em Haar-random} three-qutrit pure states, with a probability of $\approx 11.4\%$, also exhibit the same kind of transitivity property. However, at this point in writing, we still do not know whether the reduced states of any {\em three-qubit} states can ever demonstrate nonlocality transitivity, a problem we leave for future investigation.

From a foundational perspective, we may also be interested in converting examples of nonlocality transitivity for quantum states to one for {\em correlations}. Such attempts will have to overcome two major challenges. First, each party, say B, must implement the same set of measurements to exhibit the nonlocality with other parties.  Unfortunately, our construction based on multiple copies of the $W$-state marginals and the KV game does not seem to meet this requirement. Conceivably, one can use other Bell inequalities to demonstrate their nonlocality simultaneously. However, a brute force numerical optimization will be infeasible in this huge Hilbert space ($d=2^{29}\approx5.3687\times10^8$). Second, even if the local measurements for A, B, and C result in marginal correlations $\vecPAB$, $\vecPBC$, and $\vecPAC$ that are {\em each}  {\em Bell-nonlocal}, we still need $(\vecPAB, \vecPBC)$ to be constraining enough that any $\vecPAC'$  derived from compatible nonsignaling (NS) $\vecPABC'$ {\em must} also be Bell-inequality violating. Nonetheless, aside from the heuristic approaches discussed in~\cite{CorettiThesis,Coretti2011,YinMasterThesis}, we are unaware of any method to {\em systematically} search for correlations satisfying all these constraints---although the reduced states of certain Haar-random three-qutrit states do exhibit nonlocality transitivity.

Given the no-go results in~\cite{Bancal2012}, it might seem unclear why searching for an example of nonlocality transitivity in quantum theory is relevant. Two remarks are now in order. If nonlocality transitivity turned out to be {\em impossible altogether} in quantum theory, all NS theories allowing such a phenomenon would be foil theories~\cite{Foil2016}, highlighting a qualitative difference between quantum theory and its alternatives. Notice that even though we have presented examples of nonlocality transitivity of quantum states, there remains the possibility that there is {\em no} nonlocality transitivity of quantum correlations. Conversely, since nonlocality transitivity for states (correlations) subsumes the phenomenon of QNAMS (QNAMC)---which shows an unfamiliar form of nonlocality {\em polygamy}~\cite{Collins04,CKK+24,CMR24}---it is conceivable that one may derive cryptographic implications akin to those shown in~\cite{CFH21,FBL+21} or novel applications out of it. For instance, if ABC share many identical copies of $\rABC$ whose marginals exhibit nonlocality transitivity, then B could verify that AC share a nonlocal state by {\em separately} verifying the nonlocality he shares with A and C, without even needing them to measure in a spacelike separated manner.

On the other hand, since a nonlocal quantum state is necessarily steerable~\cite{WisemanPRL2007,Steering-RMP}, our construction in~\cref{Sec:NLTransitivityStates} also provides an example for steering transitivity, cf. \cref{Dfn:NLTRho} but with the word ``nonlocal'' (``nonlocality'') by ``steerable'' (``steering'') (see also~\cref{App:SteerEntTrans}).
Moreover, as we show in \cref{App:SteeringTransitivity}, even a {\em single copy} of the $W$ state marginals suffices for demonstrating this weaker form of nonlocality transitivity. From here, it seems plausible to construct an example of nonlocality transitivity in a partially device-independent setting by finding an example such that (1) A steers B, (2) C steers B, and (3) all observed measurement statistics are compatible {\em only} with A and C sharing entanglement, even without characterizing either of A's or C's measurements.

Note as well that the definition of nonlocality transitivity for quantum states can be straightforwardly generalized to other more complicated quantum networks, such as $n$ systems forming a tree graph. In fact, it is known~\cite{Tabia2022} that there are examples of $n$-qubit states such that $(n-1)$ of their qubit pair marginals exhibit entanglement transitivity for {\em all} remaining pairs. However, our constructions for nonlocality transitivity do not obviously generalize to these more general scenarios. Clearly, it will be interesting to devise tools, be it analytical or numerical, for finding examples in a general network. Of course, achieving a better understanding of the physical and practical implications of quantum nonlocality transitivity is of fundamental importance, too, that we shall pursue further in future research.

\begin{acknowledgments}
We thank Antonio Ac\'in, Nicolas Brunner, Marwan Haddara, Mu-En Liu, Hakop Pashayan, Valerio Scarani, Pavel Sekatski, Rob Spekkens, and Yujie Zhang for helpful discussions. Discussion with Pavel Sekatski, in particular, has inspired us to prove Lemma~\ref{Lem:UniqueGlobalState}, thereby leading to a stronger result of nonlocality transitivity for quantum states. KSC is grateful for the hospitality of the Institut N\'eel. This work was supported by the National Science and Technology Council, Taiwan (Grants No. 109-2112-M-006-010-MY3, 112-2628-M-006-007-MY4, 113-2917-I-006-023, 113-2918-I-006-001), the Foxconn Research Institute, Taipei, Taiwan, and in part by the Perimeter Institute for Theoretical Physics. Research at Perimeter Institute is supported by the Government of Canada through the Department of Innovation, Science, and Economic Development, and by the Province of Ontario through the Ministry of Colleges and Universities.
CYH acknowledges support from ICFOstepstone (the Marie Sk\l odowska-Curie Co-fund GA665884), the Spanish MINECO (Severo Ochoa SEV-2015-0522), the Government of Spain (FIS2020-TRANQI and Severo Ochoa CEX2019-000910-S), Fundaci\'o Cellex, Fundaci\'o Mir-Puig, Generalitat de Catalunya (SGR1381 and CERCA Programme), the ERC Advanced Grant (on grants CERQUTE and FLQuant), the AXA Chair in Quantum Information Science, the Royal Society through Enhanced Research Expenses (on grant NFQI), and the Leverhulme Trust Early Career Fellowship (on grant ``Quantum complementarity: a novel resource for quantum science and technologies'' with grant number ECF-2024-310). KSC and GNMT contributed equally to this work.
\end{acknowledgments}

\appendix

\section{Other potential candidates for nonlocality transitivity of quantum states}
\label{App:Candidates}

Here, we provide further details on why the candidate tripartite states from~\cite{scarani_gisin_2005,Collins04,BV2012}, which show, respectively, nonlocality in one, two, and three of its reduced states, are not (readily) examples exhibiting the nonlocality transitivity of quantum states. Likewise, we explain why other examples of entanglement transitivity given in~\cite{Tabia2022} do not immediately give examples of nonlocality transitivity.

\subsection{A family of qubit-qutrit-qubit states}
\label{App:SG05}

The candidate tripartite pure states of~\cite{scarani_gisin_2005} are:
\begin{equation}
	\ket{\Psi(\alpha)} = \cos\alpha\frac{\ket{021}+\ket{120}}{\sqrt{2}} +  \sin\alpha\frac{\ket{000}+\ket{111}}{\sqrt{2}},
\end{equation} 
where $\alpha\in(0,\frac{\pi}{2})$. Let $\rABC(\alpha)=\proj{\Psi(\alpha)}$, $\ket{\Psi_1(\alpha)}=\sin\alpha\ket{00}+\cos\alpha\ket{12}$, $\ket{\Psi_2(\alpha)}=\sin\alpha\ket{11}+\cos\alpha\ket{02}$, and $\Pi_{ij}:=\proj{i,j}$. As shown in~\cite{scarani_gisin_2005}, the {\em only} tripartite quantum state $\rABC'(\alpha)$, pure or mixed, compatible with the marginals
\begin{align}\label{Eq:SymmMarginals}
	\rAB(\alpha)&=\tr_\tC\left[\rABC(\alpha)\right] = \tr_\tA\left[\rABC(\alpha)\right] = \rho_\text{CB}(\alpha),\\
	&=\frac{1}{2}\proj{\Psi_1(\alpha)}+\frac{1}{2}\proj{\Psi_2(\alpha)}
\end{align} 
is $\rABC'(\alpha)=\rABC(\alpha)$. Moreover, the AC marginal 
 \begin{equation*}
 	\rAC'(\alpha)=\tr_\tB\rABC'=\cos^2\alpha\proj{\Psi^+}+\frac{\sin^2\alpha}{2}\sum_{i=0}^1\Pi_{ii}
 \end{equation*}
provably~\cite{HHH95} violates the CHSH-Bell inequality for $\cos^2\alpha>\frac{1}{\sqrt{2}}$.

However, we do {\em not} know if the AB and BC marginals for any $\alpha$ violate {\em a} Bell inequality. In particular, since $\ket{\Psi(\alpha)}$ is a symmetric extension~\cite{DPS:PRL:2002} of these marginals, we know from~\cite{Terhal03} that they cannot violate any Bell inequality with two settings on A (or C) and an arbitrary number of settings on B. We have also not found a violation of these states in the Bell scenarios where both parties perform three trichotomic measurements or four dichotomic measurements.

\subsection{A family of three-qubit states} 
\label{App:CG04}

The candidate tripartite pure states of~\cite{Collins04} are:
\begin{equation}\label{Eq:StateCG}
	\ket{\Psi(\mu)} = \mu\ket{000} + \sqrt{\frac{1-\mu^2}{2}}(\ket{110}+\ket{011}),
\end{equation} 
where we take $\mu\in[0,1]$. As with the last example, \cref{Eq:SymmMarginals} holds. Moreover, numerically, we have found that the only tripartite state compatible with these marginals is the three-qubit pure state of \cref{Eq:StateCG}. On the other hand, the AC marginal $\rAC$ is a mixture similar to \cref{Reduced_state} but with the weight of the $\ket{\Psi^+}$ given, instead, by $1-\mu^2$, i.e., 
\begin{equation}\label{Eq:W-like:Reduced}
    \rAC(\mu)= (1-\mu^2)\proj{\Psi^+}+\mu^2\proj{00}.
\end{equation}

\begin{figure}[h]
    \centering    \includegraphics[width=\columnwidth]{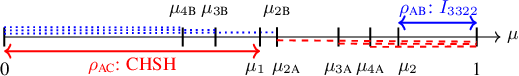}
\caption{Number line for $\mu$ showing the (non)locality of the reduced states $\rAB=\rho_\text{CB}$ and $\rAC$ of \cref{Eq:StateCG}. In the interval $\mu\in[0,\mu_1)$ with $\mu_1\approx0.5412$, $\rAC$ violates the CHSH-Bell inequality, but is symmetric-extendible (red, dashed lines) to two, three, and four A's, respectively, for $\mu$ {\em larger} than $\mu_{2\tA}\approx 0.5773,\mu_{3\tA}\approx 0.7071$, and $\mu_{4\tA}\approx0.7746$. Note that the reduced state $\tau_3$ of the three-qubit $W$ state, cf.~\cref{Reduced_state} for $n=3$, corresponds to $\mu=\frac{1}{\sqrt{3}}\approx0.5774$. Meanwhile, $\rAB$ violates the $I_{3322}$ inequality for $\mu\ge\mu_2\approx0.8343$ but is symmetric-extendible (blue, dotted lines) to two, three, and four B's, respectively, for $\mu$ {\em smaller} than $\mu_{2\tB}\approx 0.5774$, $\mu_{3\tB}\approx 0.4472$, and $\mu_{4\tB}\approx 0.3779$; $\rAB$ is also symmetric-extendible to two A's for all values of $\mu$.}
\label{Fig:NonlocalityCGState}
\end{figure}

Let $\ket{\tilde{\phi}(\mu)}:=\mu\ket{00}+ \sqrt{\frac{1-\mu^2}{2}}\ket{11}$ be a subnormalized, skewed $\ket{\Phi^+}$ state. In~\cite{Collins04}, the authors remarked that the AB marginal 
\begin{equation}
	\rAB(\mu)=\proj{\tilde{\phi}(\mu)} +  \frac{1-\mu^2}{2}\Pi_{01}
\end{equation}
violates the $I_{3322}$ Bell inequality~\cite{Collins04} when $\mu=0.852$. Moreover, due to the symmetry between A and C, if C adopts the same measurement bases as A, we can achieve a simultaneous violation for $\rBC$. Indeed, using the heuristic algorithm from~\cite{LD07}, we have found numerically that both marginals can simultaneously violate the $I_{3322}$ Bell inequality for $0.8343\lesssim\mu<1$, whereas (from~\cite{HHH95}) $\rAC$ violates the CHSH-Bell inequality only for $0\le\mu\lesssim0.5412$. However, we have not found any value of $\mu$ where all these reduced states simultaneously violate a Bell inequality. \cref{Fig:NonlocalityCGState} summarizes what we know about the nonlocality of these marginals.

\subsection{Three-qutrit state from~\cite{BV2012}} 
\label{App:BVExample}

The candidate three-qutrit pure state from~\cite{BV2012} reads as
\begin{equation}\label{Eq:psi_BV}
	\ket{\psi} = a\ket{000} + b(\ket{012} + \ket{201} + \ket{120}).
\end{equation}
After partial tracing out any of the parties, we get
\begin{equation}\label{Eq:BV-Reduced}
\begin{aligned}
	\rAB &= \rBC = \rho_\text{CA} \\
	&= (1-2b^2)\proj{\psi_\theta}+b^2(\proj{01}+\proj{20})	
\end{aligned}
\end{equation}
where $\ket{\psi_\theta} = \cos\theta\ket{00} + \sin\theta\ket{12}$ and $b^2 = \frac{\sin^2\theta}{2\sin^2\theta+1}$. For $\theta\in(0,\frac{\pi}{2})$, it was shown in~\cite{BV2012} that the marginal states of~\cref{Eq:BV-Reduced} violate the CHSH-Bell inequality.

Note that we can also recover the $\rAB$ and $\rBC$ marginals by considering a {\em different} global state:
\begin{equation}
	\vrABC = \proj{\tilde{\psi}} + b^2\proj{201}
\end{equation}
where $\ket{\tilde{\psi}} = a\ket{000} + b(\ket{012} +\ket{120})$ is a subnormalized state with norm $a^2+2b^2$. Moreover, the AC marginal of $\vrABC$ is
\begin{equation}
	\varrho_\text{AC} = a^2\proj{00} + b^2(\proj{02} + \proj{10} + \proj{21}),
\end{equation}
which is separable. Since entanglement transitivity~\cite{Tabia2022}, cf.~\cref{Dfn:EntTransitivity}, is a prerequisite for the nonlocality transitivity of quantum states (see~\cref{App:OtherDefinitions} for details), the reduced states of~\cref{Eq:psi_BV} {\em cannot} exhibit nonlocality transitivity.

\subsection{Werner states marginals} 
\label{App:EntTrans}

Apart from the Haar-random three-qudit pure states discussed in~\cref{Sec:NumExamples}, it was also shown in~\cite{Tabia2022} that if $\rAB$ and $\rBC$ are high-dimensional entangled Werner states~\cite{Werner:PRA:1989} $W^d(v)$ with $v_\text{AB}$ and $v_\text{BC}$ satisyfing some polynomial constraints, the resulting $\rAC'$ must again be entangled. However, {\em no} higher-dimensional Werner states are currently known to violate a Bell inequality, see, e.g.,~\cite{FTC+25} for a detailed discussion. Thus, these marginal density matrices are also not readily applicable to show nonlocality transitivity.

\section{Non-quantumness of the example from~\cite{Coretti2011} }\label{App:Postquantum}

Here, we provide evidence showing that the example provided by Coretti {\em et al.} (see FIG.2 of~\cite{Coretti2011}) is not quantum realizable. To this end, it suffices to show that the bipartite marginals $\vecPAB=\vecP_\text{CB}$ derived from their $\vecPABC\in\NS$ (for a Bell scenario involving four inputs for each party and two outputs for each input) are {\em outside} $\Q_1$, the level 1 {\em outer approximation} of the quantum set $\Q$ introduced by Navascu\'es-Pironio-Ac\'{\i}n (NPA) in~\cite{NPA,NPA2008}. Due to the symmetry in swapping A~$\leftrightarrow$~C, we may focus on the marginal $\vecPAB$ without loss of generality.

Let $\vecP_w$ be the uniform distribution defined in the space of $\vecPAB$, i.e., $P_w(a,b|x,y)=\frac{1}{4}$ for all $a,b\in\{1,2\}$ and all $x,y\in\{1,2,3,4\}$. Consider now the following semidefinite program (SDP) that solves for the white-noise visibility of any given  $\vecP$ with respect to $\Q_1$:
\begin{equation} \label{Eq:Visibility}
\begin{split}
   v^\star:= \max \quad & v\\
     \text{such that} \quad  v\vecP + (&1-v)\vec{P}_w \in \Q_1.
\end{split}
\end{equation}
Since $\vecP_w\in\Q\subset\Q_1$, $v=0$ is always a feasible solution to the SDP. Moreover,  if $\vecPAB\not\in\Q_1$, the visibility $v^\star$ has to be less than 1 for its convex mixture with $\vecP_w$ to lie in $\Q_1$. Conversely, if $\vecP\in\Q_1$, one finds $ v^* \geq 1$ since $ v = 1$ is always a valid solution. 

Solving~\cref{Eq:Visibility} for $\vecP=\vecPAB$ from~\cite{Coretti2011} gives $v^* \approx 0.8571$, confirming that (the marginals $\vecPAB$ from) their example is not quantum realizable. In particular, since $\Q_1$ is precisely the set of correlations satisfying the principle of {\em macroscopic locality} proposed in~\cite{Navascues:2009aa}, the marginals of the example from~\cite{Coretti2011} are incompatible with this principle. In a similar manner, one can also show that the example given in Appendix B.1 of~\cite{CorettiThesis} is not quantum realizable.

\section{Other related definitions}
\label{App:OtherDefinitions}

\subsection{Nonlocality transitivity of quantum marginal correlations}
\label{App:NTCQ}

Instead of \cref{Dfn:NTCQ}, one might also be tempted to adopt the following definition for quantum nonlocality transitivity.
\begin{definition}[Nonlocality transitivity of quantum marginals]\label{Dfn:NTCQM}
	{\em
    The pair of compatible marginals $\vecPAB$ and $\vecPBC$ exhibits {\em nonlocality transitivity of quantum marginals} if 
    \begin{enumerate}
    \item They satisfy \cref{Dfn:NTC}.
    \item They are both quantum realizable, i.e., $\vecPAB,\vecPBC\in\Q$.
    \end{enumerate}
    }
\end{definition}

In particular, one may think that if $\vecPAB$ and $\vecPBC$ are compatible, i.e., there exists $\vecPABC\in\NS$ that returns both of them as marginals via \cref{Eq:NS3}, and if these marginals are quantum realizable, then one can also find $\vecPABC\in\Q$ that returns both $\vecPAB$ and $\vecPBC$ as marginals. However, this intuition is misguided, as one can find explicit counterexample(s)~\cite{KSChen_unpublished}. 

\subsection{Steering and entanglement transitivity of quantum states}
\label{App:SteerEntTrans}

The problem of determining if compatible reduced states $\rAB$ and $\rBC$ exhibit nonlocality transitivity is an instance of a resource marginal problem~\cite{Hsieh2024resourcemarginal}. In this regard, another closely related notion of nonlocality transitivity, which can be seen as a relaxation of~\cref{Dfn:NLTRho}, can also be defined based on the phenomenon of {\em quantum steering}~\cite{Steering-RMP}. Since quantum steering is generally directional~\cite{BVQ+14}, the Definition below should be understood as being only a particular representative of an entire family of possible Definitions.
\begin{definition}[Steering transitivity of quantum states]\label{Dfn:SteeringTransitivityRho}
{\em
	The pair of compatible marginal states $\rAB$ and $\rBC$ exhibit {\em steering transitivity} from A to C if 
    \begin{enumerate}
    \item $\rAB$ is steerable from A to B. 
    \item $\rBC$ is steerable from B to C. 
    \item For all tripartite state $\rho'_\text{ABC}$ that return $\rAB$ and $\rBC$ as reduced states, the corresponding reduced state $\rho'_\text{AC}$ is steerable from A to C.
    \end{enumerate}
}
\end{definition}

The mathematical expression corresponding to \cref{Dfn:SteeringTransitivityRho} is completely analogous to \cref{Eq:NLTransitivityRho}, except that we replace $\D_\cL$ by the set of unsteerable states. 

Since only entangled quantum states can be nonlocal~\cite{Werner:PRA:1989} or steerable~\cite{WisemanPRL2007}, we see that a pre-requisite for the existence of nonlocality transitivity for quantum states is entanglement transitivity~\cite{Tabia2022}, which we recapitulate below for completeness.
\begin{definition}[Entanglement transitivity~\cite{Tabia2022}]\label{Dfn:EntTransitivity}
{\em 
	The pair of compatible marginal states $\rAB$ and $\rBC$ exhibit entanglement transitivity if 
    \begin{enumerate}
    \item They are both entangled.
    \item For all tripartite state $\rho'_\text{ABC}$ that returns $\rAB$ and $\rBC$ as reduced states, the corresponding reduced state $\rho'_\text{AC}$ is also entangled.
    \end{enumerate}
	}
\end{definition}

\begin{figure}[h!tbp]
    \centering    \includegraphics[width=0.95\columnwidth]{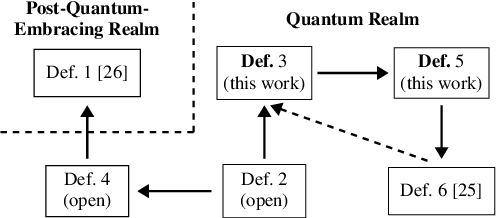}
\caption{Summary of the relationships between the various Definitions discussed in this work. The significance of the arrow is the same as that explained in~\cref{Fig:RelateDefinitions}. Incidentally, some examples satisfying~\cref{Dfn:EntTransitivity} from~\cite{Tabia2022} are also examples meeting~\cref{Dfn:NLTRho} (symbolically represented by the dashed arrow in the Figure), see~\cref{Sec:NumExamples} for details. From the arrow, we know that the example presented in~\cref{Sec:BNLTransitivity} is also an example satisfying \cref{Dfn:SteeringTransitivityRho}. However, a much simpler example can be given, see~\cref{App:SteeringTransitivity} for details.}
\label{Fig:RelateDefinitionsAll}
\end{figure}

We summarize the relationship between all these Definitions in \cref{Fig:RelateDefinitionsAll}.

\section{Khot-Vishnoi game and nonlocality from tensoring}\label{subsec:superactivation}\label{App:KV}

The Khot-Vishnoi (KV) nonlocal game~\cite{KhotVishnoi2005} is specified by two parameters: an integer $n=2^\ell$ (where $\ell \in \mathbb{N}$ is an integer) and a ``noise parameter'' $\eta \in [0, 1/2]$. For any given $n$, consider the group $G$ of $n$-bit strings with its group multiplication defined by the bitwise-XOR operation. Let $H$ be the Hadamard subgroup of $G$, whose codewords correspond to rows of a Hadamard matrix~\cite[Chap.~19]{AroraBoaz2009book}.
Now, take the quotient group $G/H$ comprised by the $\frac{2^n}{n}$ cosets $[x]$. Here $[x]$ means the coset of which $x$ is an element. By construction, each coset contains $n$ elements.
The inputs $(x,y)$ of the KV nonlocal game correspond to the cosets $[x]$ and $[y]$ while the outputs $(a,b)$ correspond, respectively, to elements of the chosen cosets.

To play the game, the referee randomly chooses a coset $[x]$ and an $n$-bit string $z$ (having a relatively low Hamming weight) such that the probability of the $i$-th bit of $z$ being $1$ is $\eta$. In each round, the referee would give input $[x]$ to Alice and $[y]=[x\oplus z]$ to Bob. Alice and Bob win the game if and only if $a\oplus b = z$. Buhrman {\em et al.}~\cite{Buhrman2012} showed that the classical winning probability in the KV game is upper bounded as $\omega_c \le n^{\frac{-\eta}{1-\eta}}$.

For a quantum strategy that outperforms this, one can first define an $n$-dimensional vector $\ket{\psi_t}$ for any $n$-bit string $t$ such that $ \ket{\psi_t}= \frac{(-1)^{t(i)}}{\sqrt{n}}\ket{i}$, where $t(i)$ is the $i$-th bit of $t$ and $\{\ket{i}\}_{i=0}^{n-1}$ is the set of computational basis vectors for $\mathbb{C}^{n}$.
For each coset $[x]$, the set of projectors $\{M_{t|x}\coloneqq\proj{\psi_t}: t\in [x]\}$ form a projective measurement since each coset is defined via the Hadamard subgroup of $G$. Applying these measurements to the $n$-dimensional maximally entangled state $\ket{\Phi_n}\coloneqq\frac{1}{\sqrt{n}}\sum_{i=0}^{n-1}\ket{i}\ket{i}$ then gives~\cite{Buhrman2012} a lower bound on the quantum winning probability $\omega_Q \ge (1- 2\eta)^2$.

For $\eta = \frac{1}{2} - \frac{1}{\ln n}$ and sufficiently large $n$, one can verify that the above lower bound on the quantum winning probability $\omega_Q\ge  \frac{4}{(\ln n)^2}$ exceeds the upper bound on the classical winning probability $\omega_c \le n^{-1 + \frac{4}{2+\ln n}}\le e^4/n$. 
It is expedient to express this in terms of the nonlocality fraction~\cite{Cavalcanti2013} (see also~\cite{Junge2011}) $\mathrm{LV}(\rho)\coloneqq \frac{\omega_Q}{\omega_c}$, where a Bell-inequality violation by $\rho$ is signified by $\mathrm{LV}(\rho)>1$.  
Thus, for $\rho=\proj{\Phi_n}$ and the measurement strategy explained above, we have 
\begin{equation}\label{Eq:LV}
	\mathrm{LV}(\rho) \ge  \frac{4n^{1-\frac{4}{2+\ln n}}}{(\ln n)^2} \ge  \frac{4n}{(\ln n)^2e^4},
\end{equation} 
which exceeds unity for sufficiently large $n$.\footnote{For the first (second) lower bound to exceed $1$, $n\in\mathbb{N}$ needs to be larger than or equal to $66$ ($541$), which corresponds to $\ell \ge 7$ ($\ell \ge 10$).}

We are now ready to recapitulate the result from~\cite{Cavalcanti2013}, showing that for any bipartite quantum state $\rho$ acting on $\mathbb{C}^d\otimes\mathbb{C}^d$ with a fully entangled fraction (FEF)~\cite{HorodeckiMP1999} larger than $\frac{1}{d}$, 
$\rho^{\otimes k}$ for a sufficiently large $k$ gives $\mathrm{LV}(\rho^{\otimes k})>1$.
To begin with, recall from~\cite{HorodeckiMP1999} the $d$-dimensional isotropic state:
\begin{equation}\label{Eq:RhoIso}
    \rho_{\mathrm{iso},d}(F) = F \proj{\Phi_d} + (1-F)\frac{\id_{d^2}-\proj{\Phi_d}}{d^2 - 1},
\end{equation}
where $\id_{d^2}$ is thd identity opeartor acting on $\mathbb{C}^d\otimes\mathbb{C}^d$ and $F=\bra{\Phi_d}\rho_{\mathrm{iso},d}(F)\ket{\Phi_d}$ is the so-called singlet fraction~\cite{HHH99Teleportation} of $\rho_{\mathrm{iso},d}(F)$, which coincides with its FEF whenever $F\ge\frac{1}{d^2}$. In general, for any given $\rho$, its FEF is determined as $F_\rho\coloneqq\max_{\Phi'_d} \bra{\Phi'_d}\rho\ket{\Phi'_d}$ with the maximum taken over all $d$-dimensional maximally entangled state $\ket{\Phi'_d}=\id\otimes V\,\ket{\Phi_d}$ and $V$ is a unitary operator~\cite{HorodeckiMP1999}.

Notice that $k$ copies of the isotropic can be written as
\begin{equation}
    \rho_{\mathrm{iso},d}^{\otimes k} = F^k \proj{\Phi_{d^k}} + \cdots,
\end{equation}
which is a convex mixture of the $d^k$-dimensional maximally entangled state $\ket{\Phi_{d^k}}=\ket{\Phi_{d}}^{\otimes k}$ with other noise terms.
By considering only the contribution from this first term in $\omega_Q$ and setting $n=d^k$, we get the lower bound
\begin{equation}\label{eq:kcopyNLfrac}
	\mathrm{LV}(\rho_{\mathrm{iso},d}^{\otimes k})\ge F^k \mathrm{LV}(\proj{\Phi_{d^k}}) \ge \frac{4}{e^4}\frac{(Fd)^k}{(k \ln d)^2}, 
\end{equation}
which, for $F>\frac{1}{d}$, will exceed unity for sufficiently large $k$.

Next, recall from~\cite{HorodeckiMP1999} that any $\rho$ can be depolarized into an isotropic state by the $U\otimes \bar{U}$ twirling (here, $\bar{U}$ is the complex conjugate of an arbitrary unitary $U$) while leaving its singlet fraction unchanged. Using this observation and some convexity argument, it can be shown that via the KV nonlocal game,\footnote{\label{fn:FEFvsF} If the initial state $\rho$ has an FEF $F_\rho=\bra{\Phi'_d}\rho\ket{\Phi'_d}$ {\em larger} than its singlet fraction, then one should first perform the local unitary transformation $\id\otimes V^\dag$ on $\rho$, where $\ket{\Phi'_d}=\id\otimes V\,\ket{\Phi_d}$. Then, a follow-up $U\otimes \bar{U}$ twirling will convert $(\id\otimes V^\dag)\rho(\id\otimes V)$ with a singlet fraction of $F_\rho$ to $\rho_{\mathrm{iso},d}(F_\rho)$.}
\begin{equation}\label{Eq:rho-rhoF}
	\mathrm{LV}^*(\rho^{\otimes k})\ge \mathrm{LV}[\rho_{\mathrm{iso},d}^{\otimes k}(F_\rho)]
\end{equation}
where we use $\mathrm{LV}^*(\rho^{\otimes k})$ to represent the maximal quantum winning probability for the KV game achievable using $\rho^{\otimes k}$. From \cref{Eq:rho-rhoF,eq:kcopyNLfrac}, we see that $\mathrm{LV}^*(\rho^{\otimes k})>1$  if $F_\rho>\frac{1}{d}$. See~\cite{Cavalcanti2013} for details. 

\section{Proofs leading to our analytic examples of nonlocality transitivity for quantum states}

\subsection{Proof of  Lemma~\ref{Lem:UniqueGlobalState}}
\label{App:ProofUniqueness}

In the following, we give the proof of Lemma~\ref{Lem:UniqueGlobalState}, which concerns the uniqueness of a global state compatible with the specification of $(n-1)$ two-body marginals of $\ket{W_n}^{\otimes k}$ for any integer $k\ge 1$.
\begin{proof}
Let $\{\ket{i}: i=0,1,\ldots,d-1\}$ be the standard basis for each party, where $d=2^k$. Mathematically, each local $d$-dimensional Hilbert space is isomorphic to a $k$-qubit state space. Therefore, we may also express each local basis state as a $k$-qubit basis state where $i$ is expressed in its binary representation from right to left:
\begin{align}
        \ket{0}&= \ket{00\cdots 0}, &
    \ket{1} &= \ket{10\cdots 0}, \nonumber \\
    \ket{2} &= \ket{01\cdots 0}, & \ket{3} &= \ket{11\cdots 0}, \nonumber \\
    &\cdots & \ket{2^k-1} &= \ket{11\cdots 1}.
\end{align}
At this point, this alternative representation using $k$ qubits is merely a mathematical convenience, which does not {\em a priori} require one to assume that each party has access to $k$ two-level systems. However, it facilitates our reference to the bit value of the $k$-th virtual qubit (hereafter {\em vbit}) for each party, which simplifies our discussion (see \cref{Fig:k_Copies_WState_Marginal}).

\begin{figure}[h!tbp]
    \centering    \includegraphics[width=0.88\columnwidth]{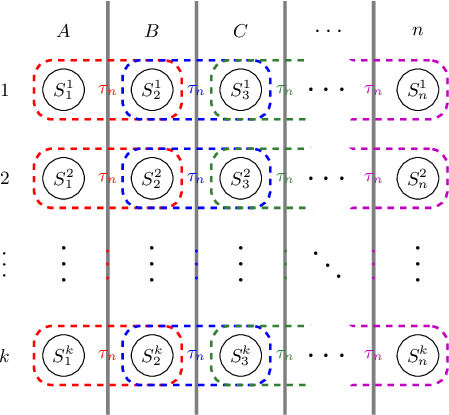}
\caption{Schematic representation of the $2^{kn}$-dimensional Hilbert space shared by $n$ parties each holding $k$ virtual qubits (vbits). Here $S_j^m$ refers to the $m$-th vbit of the $j$-th party. The premise of Lemma~\ref{Lem:UniqueGlobalState} effectively demands that all pairs of ``neighboring'' vbits (enclosed by a dashed oval) must be in the state $\tau_n$, thus leading to \cref{Eq:alpha-constr1}.
}
\label{Fig:k_Copies_WState_Marginal}
\end{figure}

Now consider an arbitrary $n$-partite global state $\rho'_S$ acting on $(\mathbb{C}^{d})^{\otimes n}$ with $d = 2^k$. Writing $\rho'_S$ in its spectral decomposition with non-vanishing eigenvalues $c_\ell> 0$ gives:
\begin{subequations}\label{Eq:rhoS}
\begin{gather}
    \rho'_S = \sum_\ell c_\ell \proj{\Psi_\ell},\\
    \ket{\Psi_\ell} \coloneqq \sum_{i_1^1,i_2^1,\dots,i_1^k,\dots,i_{n}^k} \alpha^{(\ell)}_{i_1^1,i_2^1,\dots,i_{n}^k} \ket{i_1^1,i_2^1,\dots,i_1^k,\dots,i_{n}^k},
\end{gather}
\end{subequations}
where $ \ket{\Psi_\ell}$ is an eigenket of $\rho'_S$, 
$i_j^m\in\{0,1\}$ is the bit value associated with the $m$-th vbit of the $j$-th party. 
We can also define partial traces for the vbits so that the total trace becomes:
\begin{equation}
    \tr = \tr_{S_1^1}\tr_{S_1^2}\cdots\tr_{S_1^k}\tr_{S_2^1}\cdots\tr_{S_n^k},
\end{equation}
where the index $S_j^m$ refers to the $m$-th vbit of the $j$-th party.

Let $\ket{xy}_{S_{i,j}^m} \coloneqq \ket{x}_{S_i^m}\ket{y}_{S_j^m}$ denote the product state where the $m$-th vbit of parties $i$ and $j$ are, respectively, $\ket{x}$ and $\ket{y}$. From the premise of the Lemma, we see that these vbits of $S^m_{i,j}$ (for {\em any} $m=1,2,\cdots, k$) must be in a two-qubit ``reduced state'' of $\rho'_S$ consistent with $\tau_n$ [cf. \cref{Reduced_state}], i.e.,
\begin{equation}\label{Eq:TwoVbit}
	\tr_{S\backslash S^m_{i,j}}\rho'_S = \tau_n\quad\forall~i,j\neq i\in\{1,2,\cdots,n\},
\end{equation}
where $\tr_{S\backslash s}$ refers to a partial trace over all but the $s$-th vbits. For \cref{Eq:TwoVbit} to hold, the reduced two-vbit state, which equals $\tau_n$, cf.~\cref{Reduced_state}, must 
not have support on the subspace orthogonal to $\tau_n$, i.e., can have no contribution from $\mathrm{span}\{\ket{11},\ket{\Psi^{-}}\}$, where $\ket{\Psi^-}$ is the two-qubit singlet state.
Specifically, applying the above observation to the $S^1_{j,j+1}$ vbits of $\rho'_S$ for $j\in\{1,2,\cdots,n-1\}$ gives:\footnote{It is understood that for $j=n-1$, the index $i^m_{j+2}$ is absent.} 
\begin{equation}\label{Eq:ZeroConstr}
\begin{split}
        0 &=\bra{11}\tau_n\ket{11}_{S_{j,j+1}^1} = \bra{11}\tr_{S\backslash S^1_{j,j+1}}\rho'_S\ket{11}_{S_{j,j+1}^1} \\
        &=\bra{11}\tr_{S\backslash S^1_{j,j+1}}\left(\sum_{\ell}c_{\ell}\proj{\Psi_{\ell}}\right)\ket{11}_{S_{j,j+1}^1} \\
   \Rightarrow    0 &= \sum_{\ell}c_{\ell}\sum_{i_1^1,\dots,i_{j-1}^1,i_{j+2}^1,\dots,i_{n}^k} |\alpha^{(\ell)}_{i_1^1,\dots,i_{j-1}^1,1,1,i_{j+2}^1,\cdots,i_{n}^k}|^2. 
\end{split}
\end{equation}
Remembering that $c_\ell>0$ for all $\ell$, \cref{Eq:ZeroConstr} implies that
\begin{equation}\label{Eq:alpha-constr1}
	\alpha^{(\ell)}_{i_1^1,\ldots,i_{j-1}^1,1,1,i_{j+2}^1,\ldots, i_{n}^k}=0\quad\forall~i_1^1,\dots,i_{j-1}^1,i_{j+2}^1,\cdots,i_{n}^k.
\end{equation}
Since the same conclusion holds for all \mbox{$j\in\{1,2,\cdots,n-1\}$,} the amplitude $\alpha^{(\ell)}_{i_1^1,\ldots, i_{n}^k}$ vanishes whenever two adjacent indices $i^m_j,i^m_{j+1}$ both take the value $1$.
Similarly, from $\bra{\Psi^-}\tau_n\ket{\Psi^-}_{S_{j,j+1}^1}=0$, we see that
\begin{gather}
        \left(\frac{\bra{01}-\bra{10}}{\sqrt{2}}\right)\tr_{S\backslash S^1_{j,j+1}}\rho'_S\left(\frac{\ket{01}-\ket{10}}{\sqrt{2}}\right)_{S_{j,j+1}^1}=0,\nonumber\\
        \implies \alpha^{(\ell)}_{i_1^1,\dots,i_{j-1}^1,1,0,i_{j+2}^1,\cdots,i_{n}^k}=\alpha^{(\ell)}_{i_1^1,\dots,i_{j-1}^1,0,1,i_{j+2}^1,\cdots,i_{n}^k} \nonumber\\
        \forall~i_1^1,\dots,i_{j-1}^1,i_{j+2}^1,\cdots,i_{n}^k.\qquad\qquad
        \label{Eq:alpha-constr2}
\end{gather}
Putting the observations from \cref{Eq:alpha-constr1,Eq:alpha-constr2} together, we see that $\ket{\Psi_\ell}$ must have {\em no} contribution from basis states where two or more of the $m$-th vbits are in the state $\ket{1}$.
Hence, the eigenket $\ket{\Psi_\ell}$ must take the form
\begin{equation}
\begin{aligned}
	\ket{\Psi_\ell}=\sum_{i_1^2,\dots,i_{n}^k}&~~\alpha^{(\ell)}_{\vec{0}_n,i_{1}^2,\cdots,i_{n}^k}\ket{0}^{\otimes n}\ket{i^2_1,\cdots,i^k_n}\\
	+&~~\alpha^{(\ell)}_{1,\vec{0}_{n-1},i_{1}^2,\cdots,i_{n}^k}\ket{\widetilde{W}_n}\ket{i^2_1,\cdots,i^k_n}
\end{aligned}
\end{equation}
where $\vec{0}_n$ represents an $n$-bit string of zeros and \mbox{$\ket{\widetilde{W}_n}\coloneqq\sqrt{n}\ket{W_n}$.} Continuing the same analysis for the vbits of $S^m_{j,j+1}$ (with $m=2,3,\cdots,k$) gives
\begin{equation}\label{Eq:Psi_l2}
	\ket{\Psi_\ell}=\sum_{\iota^1,\iota^2,\dots,\iota^k}~\beta^{(\ell)}_{\iota^1,\iota^2,\cdots,\iota^k}\ket{\iota^1,\iota^2,\cdots,\iota^k},
\end{equation}
where each $\ket{\iota^m}$ is either the $n$-qubit product state $\ket{0}^{\otimes n}$ or the $n$-qubit $W$-state $\ket{W_n}$.

Next, from \cref{Reduced_state,Eq:rhoS,Eq:Psi_l2,Eq:TwoVbit} and by considering, say, the $S^1_{1,2}$ vbits, we get
\begin{align}
	&\bra{\Psi^+}\tau_n\ket{\Psi^+}_{S_{1,2}^1} = \bra{\Psi^+}\tr_{S\backslash S^1_{1,2}}\rho'_S\ket{\Psi^+}_{S_{1,2}^1} \nonumber\\
	\Rightarrow~ & \frac{2}{n} =\sum_\ell c_\ell \bra{\Psi^+}\left(\tr_{S\backslash S^1_{1,2}}\proj{\Psi_\ell}\right)\ket{\Psi^+}_{S_{1,2}^1} \nonumber\\	
	 \Rightarrow~ & \frac{2}{n} =\sum_\ell c_\ell \sum_{\iota^2,\dots,\iota^k} \left|\beta^{(\ell)}_{W_n,\iota^2,\cdots,\iota^k}\right|^2\left|\bra{W_n}\left(\ket{\Psi^+}\ket{0}^{\otimes(n-2)}\right)\right|^2 \nonumber\\	
	 \Rightarrow~ & \frac{2}{n} =\sum_\ell c_\ell \sum_{\iota^2,\dots,\iota^k} \left|\beta^{(\ell)}_{W_n,\iota^2,\cdots,\iota^k}\right|^2\frac{2}{n} \nonumber\\	
	\Rightarrow~ & 
	\sum_{\iota^2,\dots,\iota^k} \left|\beta^{(\ell)}_{W_n,\iota^2,\cdots,\iota^k}\right|^2=1\quad\forall\,\,\ell, 
	\label{Eq:beta-constr}
\end{align}
where the last equality follows from $\sum_\ell c_\ell =1$ and the fact that $c_\ell>0$, cf.~\cref{Eq:rhoS}. Then, by comparing \cref{Eq:beta-constr} with the normalization of $\ket{\Psi_\ell}$ in terms of its amplitude, we see that $\beta^{(\ell)}_{\vec{0}_n,\iota^2,\cdots,\iota^k}$ {\em must vanish} for all values of $\iota_2,\iota_3,\cdots,\iota_k$. By repeating similar arguments for the two vbits of $S^m_{1,2}$ for $m=2,3,\cdots,k$ eventually leads to the conclusion that for all $\ell$, $\proj{\Psi_\ell}=\proj{W_n}$, and hence 
\begin{equation}
    \rho'_S = \proj{W_n}^{\otimes k},
\end{equation}
which concludes our proof of uniqueness.

Note that in the proof above, instead of considering the \mbox{$(n-1)$} adjacent pairs from $S^m_{i,j}$ for any given $m$, which leads us to \cref{Eq:alpha-constr1,Eq:alpha-constr2}, we could just as well consider any $(n-1)$ pairs such that when these $n$ nodes are seen as the vertices of an $n$-vertex graph, the edges correspond to the pairs that do not lead to any cycle and that the graph is connected. In other words, to have a unique global state, we only need to specify $(n-1)$ bipartite marginals where these marginals correspond to the edges forming a tree graph (see~\cite{Tabia2022}).
\end{proof}

\subsection{Proof of \cref{Result:NonlocalityTransitivity}}\label{App:ProofNLTransitivityRho}

\begin{proof}
Consider the two-qubit reduced state of a three-qubit $W$ state $\ket{W_3}$, cf. \cref{Reduced_state} with $n=3$,
\begin{equation}\label{eq.3qubitWmarg}
    \tau_3 = \frac{2}{3}\proj{\Psi^+} + \frac{1}{3}\proj{00}.
\end{equation}
Let $\rAB = \rBC = \tau_3^{\otimes k}$, i.e., the $k$ copies of $\tau_3$ for some $k\in\mathbb{N}$. 
 
From Lemma~\ref{Lem:UniqueGlobalState}, we know that the only tripartite global state compatible with these marginals is \mbox{$\rABC'=(\proj{W_3})^{\otimes k}$,} which means that \mbox{$\rAC'=\tr_\tB\rABC' = (\tau_3)^{\otimes k}$.} Clearly, the FEF of $\tau_3$ is its overlap with $\ket{\Psi^+}$, i.e., $F_{\tau_3}=\frac{2}{3}>\frac{1}{2}$.
It then follows from~\cref{eq:kcopyNLfrac,Eq:rho-rhoF} that
\begin{equation}
    \mathrm{LV}^*(\tau_3^{\otimes k}) \ge  \frac{4}{e^4}\frac{(4/3)^k}{(k \ln 2)^2}>1,
\end{equation}
where the last inequality holds when $k\ge 31$. If, instead, we use the original, tighter lower bound on $\mathrm{LV}(\ket{\Psi_{2^k}})$, i.e., the first inequality of~\cref{Eq:LV}, then we have
\begin{equation}\label{Eq:LVtighter}
    \mathrm{LV}^*(\tau_3^{\otimes k}) \ge  \frac{4}{(k\ln 2)^2}\times2^{k(1-\frac{4}{2+k\ln 2})}\times \left(\frac{2}{3}\right)^k,
\end{equation}
which exceeds unity when $k\ge 29$. Therefore, by taking $29$ or more copies of $\tau_3$ as $\rAB$ and $\rBC$, these bipartite states 
\begin{enumerate}
\item[(1)] are nonlocal as they can give a winning probability of the KV nonlocal game better than any classical strategy and 
\item[(2)] together force any compatible $\rAC'$ to take the same form, and hence also nonlocal. 
\end{enumerate}
Consequently, these reduced states of $\ket{W_3}^{\otimes k}$ exhibit nonlocality transitivity for quantum states, according~\cref{Dfn:NLTRho}. See~\cref{Fig:ProofIllustration} for a schematic illustration.
\end{proof}

\section{Steering transitivity of quantum states}
\label{App:SteeringTransitivity}

Given that the set of Bell-nonlocal quantum states strictly contains the set of steerable quantum states~\cite{WisemanPRL2007}, the examples presented in~\cref{Sec:BNLTransitivity,Sec:NumExamples} are also examples exhibiting steering transitivity of quantum states, cf.~\cref{Dfn:SteeringTransitivityRho}. However, this difference in the two notions of nonlocality also allows one to identify a simpler example of steering transitivity.

Before giving this simpler example, we shall first recall the following Lemma from~\cite{Quintino2016} (see also Section III A of~\cite{Hsieh2016}), which we also provide a proof below for ease of reference.
\begin{lemma}[Sufficiency for steerability~\cite{Quintino2016}]\label{Lem:SuffFd}
	Let \mbox{$H_d\coloneqq\sum_{n=1}^d \frac{1}{n}$} be the Harmonic series. Any state $\rho$ acting on $\mathbb{C}^d\otimes\mathbb{C}^d$ and having a fully-entangled fraction (FEF) $F_\rho> \mathcal{F}^\text{steer}_d\coloneqq\frac{d+1}{d^2}H_d - \frac{1}{d}$ is steerable.
\end{lemma}
\begin{proof}
Let us start by noting that the isotropic state of \cref{Eq:RhoIso} is also commonly written as 
\begin{equation}\label{Eq:Iso2}
	\rho_{\mathrm{iso},d}(p) = p\proj{\Phi_d}+(1-p)\frac{\id_{d^2}}{d^2}.
\end{equation}
Moreover, $\rho_{\mathrm{iso},d}(p)$ is known~\cite{WisemanPRL2007,JWD:PRA:2007} to be steerable via projective measurements for $p>\frac{H_d-1}{d-1}$. By comparing \cref{Eq:Iso2} and \cref{Eq:RhoIso}, one can verify that $\rho_{\mathrm{iso},d}(p)$ has a singlet fraction $F=p+\frac{1-p}{d^2}$. This means that the isotropic state is steerable whenever $F>\mathcal{F}^\text{steer}_d\coloneqq\frac{d+1}{d^2}H_d - \frac{1}{d}$. 

Again, recall from~\cite{HorodeckiMP1999} that any bipartite $\rho$ can be converted into an isotropic state by the $(U\otimes \bar{U})$-twirling operation. Since this operation is a convex mixture of local operations, it cannot make an unsteerable $\rho$ steerable. Together with the facts that 
\begin{enumerate}
\item[(1)] $(U\otimes \bar{U})$-twirling leaves the singlet fraction of a state $\rho$ unchanged, 
\item[(2)] if $\rho$ has an FEF $F_\rho$ larger than $\mathcal{F}^\text{steer}_d$, it can be transformed by a local unitary transformation to a state having a singlet fraction equals to $F_\rho$ (see~\cref{fn:FEFvsF}), 
\end{enumerate}
then any $\rho$ with an FEF $F_\rho$ larger than $\frac{d+1}{d^2}H_d - \frac{1}{d}$ must also be steerable as claimed.
\end{proof}

In the case of $d=2$, we have $\mathcal{F}^\text{steer}_d=\frac{5}{8}=0.625$. Note that the recent result from~\cite{zhang2024exact,Renner2024PRL} implies that this threshold $\mathcal{F}^\text{steer}_2$ cannot be improved any further. However, since $\tau_3$, the two-qubit reduced state of the three-qubit $W$ state $\ket{W_3}$, has an FEF equals to $\frac{2}{3}>\frac{5}{8}$, then together with Lemma~\ref{Lem:UniqueGlobalState} for $k=1$, we thus arrive at the following corollary.
\begin{corollary}\label{Prop:SteerTransRho3}
	Marginals of the three-qubit $W$ state, i.e., $\sigma_\text{AB}=\sigma_\text{BC}=\tau_3$, exhibit the transitivity of steerability.
\end{corollary}


\begin{thebibliography}{87}%
\makeatletter
\providecommand \@ifxundefined [1]{%
 \@ifx{#1\undefined}
}%
\providecommand \@ifnum [1]{%
 \ifnum #1\expandafter \@firstoftwo
 \else \expandafter \@secondoftwo
 \fi
}%
\providecommand \@ifx [1]{%
 \ifx #1\expandafter \@firstoftwo
 \else \expandafter \@secondoftwo
 \fi
}%
\providecommand \natexlab [1]{#1}%
\providecommand \enquote  [1]{``#1''}%
\providecommand \bibnamefont  [1]{#1}%
\providecommand \bibfnamefont [1]{#1}%
\providecommand \citenamefont [1]{#1}%
\providecommand \href@noop [0]{\@secondoftwo}%
\providecommand \href [0]{\begingroup \@sanitize@url \@href}%
\providecommand \@href[1]{\@@startlink{#1}\@@href}%
\providecommand \@@href[1]{\endgroup#1\@@endlink}%
\providecommand \@sanitize@url [0]{\catcode `\\12\catcode `\$12\catcode
  `\&12\catcode `\#12\catcode `\^12\catcode `\_12\catcode `\%12\relax}%
\providecommand \@@startlink[1]{}%
\providecommand \@@endlink[0]{}%
\providecommand \url  [0]{\begingroup\@sanitize@url \@url }%
\providecommand \@url [1]{\endgroup\@href {#1}{\urlprefix }}%
\providecommand \urlprefix  [0]{URL }%
\providecommand \Eprint [0]{\href }%
\providecommand \doibase [0]{http://dx.doi.org/}%
\providecommand \selectlanguage [0]{\@gobble}%
\providecommand \bibinfo  [0]{\@secondoftwo}%
\providecommand \bibfield  [0]{\@secondoftwo}%
\providecommand \translation [1]{[#1]}%
\providecommand \BibitemOpen [0]{}%
\providecommand \bibitemStop [0]{}%
\providecommand \bibitemNoStop [0]{.\EOS\space}%
\providecommand \EOS [0]{\spacefactor3000\relax}%
\providecommand \BibitemShut  [1]{\csname bibitem#1\endcsname}%
\let\auto@bib@innerbib\@empty
\bibitem [{\citenamefont {Horodecki}\ \emph {et~al.}(2009)\citenamefont
  {Horodecki}, \citenamefont {Horodecki}, \citenamefont {Horodecki},\ and\
  \citenamefont {Horodecki}}]{Ent-RMP}%
  \BibitemOpen
  \bibfield  {author} {\bibinfo {author} {\bibfnamefont {R.}~\bibnamefont
  {Horodecki}}, \bibinfo {author} {\bibfnamefont {P.}~\bibnamefont
  {Horodecki}}, \bibinfo {author} {\bibfnamefont {M.}~\bibnamefont
  {Horodecki}}, \ and\ \bibinfo {author} {\bibfnamefont {K.}~\bibnamefont
  {Horodecki}},\ }\href {\doibase 10.1103/RevModPhys.81.865} {\bibfield
  {journal} {\bibinfo  {journal} {Rev. Mod. Phys.}\ }\textbf {\bibinfo {volume}
  {81}},\ \bibinfo {pages} {865} (\bibinfo {year} {2009})}\BibitemShut
  {NoStop}%
\bibitem [{\citenamefont {Brunner}\ \emph {et~al.}(2014)\citenamefont
  {Brunner}, \citenamefont {Cavalcanti}, \citenamefont {Pironio}, \citenamefont
  {Scarani},\ and\ \citenamefont {Wehner}}]{Bell-RMP}%
  \BibitemOpen
  \bibfield  {author} {\bibinfo {author} {\bibfnamefont {N.}~\bibnamefont
  {Brunner}}, \bibinfo {author} {\bibfnamefont {D.}~\bibnamefont {Cavalcanti}},
  \bibinfo {author} {\bibfnamefont {S.}~\bibnamefont {Pironio}}, \bibinfo
  {author} {\bibfnamefont {V.}~\bibnamefont {Scarani}}, \ and\ \bibinfo
  {author} {\bibfnamefont {S.}~\bibnamefont {Wehner}},\ }\href {\doibase
  10.1103/RevModPhys.86.419} {\bibfield  {journal} {\bibinfo  {journal} {Rev.
  Mod. Phys.}\ }\textbf {\bibinfo {volume} {86}},\ \bibinfo {pages} {419}
  (\bibinfo {year} {2014})}\BibitemShut {NoStop}%
\bibitem [{\citenamefont {Einstein}\ \emph {et~al.}(1971)\citenamefont
  {Einstein}, \citenamefont {Born},\ and\ \citenamefont
  {Born}}]{Einstein1971TheBL}%
  \BibitemOpen
  \bibfield  {author} {\bibinfo {author} {\bibfnamefont {A.~B.}\ \bibnamefont
  {Einstein}}, \bibinfo {author} {\bibfnamefont {M.}~\bibnamefont {Born}}, \
  and\ \bibinfo {author} {\bibfnamefont {H.}~\bibnamefont {Born}},\ }\href@noop
  {} {\emph {\bibinfo {title} {The {Born-Einstein} letters: Correspondence
  between {Albert Einstein} and {Max} and {Hedwig Born} from 1916-1955, with
  commentaries by {Max Born}}}}\ (\bibinfo  {publisher} {Macmillan},\ \bibinfo
  {year} {1971})\BibitemShut {NoStop}%
\bibitem [{\citenamefont {Bell}(1964)}]{Bell64}%
  \BibitemOpen
  \bibfield  {author} {\bibinfo {author} {\bibfnamefont {J.~S.}\ \bibnamefont
  {Bell}},\ }\href {https://cds.cern.ch/record/111654} {\bibfield  {journal}
  {\bibinfo  {journal} {Physics}\ }\textbf {\bibinfo {volume} {1}},\ \bibinfo
  {pages} {195} (\bibinfo {year} {1964})}\BibitemShut {NoStop}%
\bibitem [{\citenamefont {Bell}(2004)}]{Bell04}%
  \BibitemOpen
  \bibfield  {author} {\bibinfo {author} {\bibfnamefont {J.~S.}\ \bibnamefont
  {Bell}},\ }\href {\doibase 10.1017/CBO9780511815676} {\emph {\bibinfo {title}
  {Speakable and Unspeakable in Quantum Mechanics: Collected Papers on Quantum
  Philosophy}}},\ \bibinfo {edition} {2nd}\ ed.\ (\bibinfo  {publisher}
  {Cambridge University Press},\ \bibinfo {year} {2004})\BibitemShut {NoStop}%
\bibitem [{\citenamefont {Vidal}(2003)}]{VidalPRL03}%
  \BibitemOpen
  \bibfield  {author} {\bibinfo {author} {\bibfnamefont {G.}~\bibnamefont
  {Vidal}},\ }\href {\doibase 10.1103/PhysRevLett.91.147902} {\bibfield
  {journal} {\bibinfo  {journal} {Phys. Rev. Lett.}\ }\textbf {\bibinfo
  {volume} {91}},\ \bibinfo {pages} {147902} (\bibinfo {year}
  {2003})}\BibitemShut {NoStop}%
\bibitem [{\citenamefont {Jozsa}\ and\ \citenamefont
  {Linden}(2003)}]{Jozsa:2003wj}%
  \BibitemOpen
  \bibfield  {author} {\bibinfo {author} {\bibfnamefont {R.}~\bibnamefont
  {Jozsa}}\ and\ \bibinfo {author} {\bibfnamefont {N.}~\bibnamefont {Linden}},\
  }\href {\doibase 10.1098/rspa.2002.1097} {\bibfield  {journal} {\bibinfo
  {journal} {Proc. R. Soc. Lond. A.}\ }\textbf {\bibinfo {volume} {459}},\
  \bibinfo {pages} {2011} (\bibinfo {year} {2003})}\BibitemShut {NoStop}%
\bibitem [{\citenamefont {Ekert}(1991)}]{Ekert91}%
  \BibitemOpen
  \bibfield  {author} {\bibinfo {author} {\bibfnamefont {A.~K.}\ \bibnamefont
  {Ekert}},\ }\href {\doibase 10.1103/PhysRevLett.67.661} {\bibfield  {journal}
  {\bibinfo  {journal} {Phys. Rev. Lett.}\ }\textbf {\bibinfo {volume} {67}},\
  \bibinfo {pages} {661} (\bibinfo {year} {1991})}\BibitemShut {NoStop}%
\bibitem [{\citenamefont {Ac\'{\i}n}\ \emph {et~al.}(2007)\citenamefont
  {Ac\'{\i}n}, \citenamefont {Brunner}, \citenamefont {Gisin}, \citenamefont
  {Massar}, \citenamefont {Pironio},\ and\ \citenamefont {Scarani}}]{Acin07}%
  \BibitemOpen
  \bibfield  {author} {\bibinfo {author} {\bibfnamefont {A.}~\bibnamefont
  {Ac\'{\i}n}}, \bibinfo {author} {\bibfnamefont {N.}~\bibnamefont {Brunner}},
  \bibinfo {author} {\bibfnamefont {N.}~\bibnamefont {Gisin}}, \bibinfo
  {author} {\bibfnamefont {S.}~\bibnamefont {Massar}}, \bibinfo {author}
  {\bibfnamefont {S.}~\bibnamefont {Pironio}}, \ and\ \bibinfo {author}
  {\bibfnamefont {V.}~\bibnamefont {Scarani}},\ }\href {\doibase
  10.1103/PhysRevLett.98.230501} {\bibfield  {journal} {\bibinfo  {journal}
  {Phys. Rev. Lett.}\ }\textbf {\bibinfo {volume} {98}},\ \bibinfo {pages}
  {230501} (\bibinfo {year} {2007})}\BibitemShut {NoStop}%
\bibitem [{\citenamefont {Vazirani}\ and\ \citenamefont
  {Vidick}(2014)}]{Vazirani14}%
  \BibitemOpen
  \bibfield  {author} {\bibinfo {author} {\bibfnamefont {U.}~\bibnamefont
  {Vazirani}}\ and\ \bibinfo {author} {\bibfnamefont {T.}~\bibnamefont
  {Vidick}},\ }\href {\doibase 10.1103/PhysRevLett.113.140501} {\bibfield
  {journal} {\bibinfo  {journal} {Phys. Rev. Lett.}\ }\textbf {\bibinfo
  {volume} {113}},\ \bibinfo {pages} {140501} (\bibinfo {year}
  {2014})}\BibitemShut {NoStop}%
\bibitem [{\citenamefont {Coffman}\ \emph {et~al.}(2000)\citenamefont
  {Coffman}, \citenamefont {Kundu},\ and\ \citenamefont
  {Wootters}}]{Coffman2000PRA}%
  \BibitemOpen
  \bibfield  {author} {\bibinfo {author} {\bibfnamefont {V.}~\bibnamefont
  {Coffman}}, \bibinfo {author} {\bibfnamefont {J.}~\bibnamefont {Kundu}}, \
  and\ \bibinfo {author} {\bibfnamefont {W.~K.}\ \bibnamefont {Wootters}},\
  }\href {\doibase 10.1103/PhysRevA.61.052306} {\bibfield  {journal} {\bibinfo
  {journal} {Phys. Rev. A}\ }\textbf {\bibinfo {volume} {61}},\ \bibinfo
  {pages} {052306} (\bibinfo {year} {2000})}\BibitemShut {NoStop}%
\bibitem [{\citenamefont {Doherty}(2014)}]{Doherty:2014aa}%
  \BibitemOpen
  \bibfield  {author} {\bibinfo {author} {\bibfnamefont {A.~C.}\ \bibnamefont
  {Doherty}},\ }\href {\doibase 10.1088/1751-8113/47/42/424004} {\bibfield
  {journal} {\bibinfo  {journal} {J. Phys. A: Math. Theor.}\ }\textbf {\bibinfo
  {volume} {47}},\ \bibinfo {pages} {424004} (\bibinfo {year}
  {2014})}\BibitemShut {NoStop}%
\bibitem [{\citenamefont {Popescu}\ and\ \citenamefont
  {Rohrlich}(1994)}]{PopescuFP94}%
  \BibitemOpen
  \bibfield  {author} {\bibinfo {author} {\bibfnamefont {S.}~\bibnamefont
  {Popescu}}\ and\ \bibinfo {author} {\bibfnamefont {D.}~\bibnamefont
  {Rohrlich}},\ }\href {\doibase 10.1007/BF02058098} {\bibfield  {journal}
  {\bibinfo  {journal} {Found. Phys.}\ }\textbf {\bibinfo {volume} {24}},\
  \bibinfo {pages} {379} (\bibinfo {year} {1994})}\BibitemShut {NoStop}%
\bibitem [{\citenamefont {Barrett}\ \emph {et~al.}(2005)\citenamefont
  {Barrett}, \citenamefont {Linden}, \citenamefont {Massar}, \citenamefont
  {Pironio}, \citenamefont {Popescu},\ and\ \citenamefont
  {Roberts}}]{BarrettPRA05}%
  \BibitemOpen
  \bibfield  {author} {\bibinfo {author} {\bibfnamefont {J.}~\bibnamefont
  {Barrett}}, \bibinfo {author} {\bibfnamefont {N.}~\bibnamefont {Linden}},
  \bibinfo {author} {\bibfnamefont {S.}~\bibnamefont {Massar}}, \bibinfo
  {author} {\bibfnamefont {S.}~\bibnamefont {Pironio}}, \bibinfo {author}
  {\bibfnamefont {S.}~\bibnamefont {Popescu}}, \ and\ \bibinfo {author}
  {\bibfnamefont {D.}~\bibnamefont {Roberts}},\ }\href {\doibase
  10.1103/PhysRevA.71.022101} {\bibfield  {journal} {\bibinfo  {journal} {Phys.
  Rev. A}\ }\textbf {\bibinfo {volume} {71}},\ \bibinfo {pages} {022101}
  (\bibinfo {year} {2005})}\BibitemShut {NoStop}%
\bibitem [{\citenamefont {Masanes}\ \emph {et~al.}(2006)\citenamefont
  {Masanes}, \citenamefont {Ac\'in},\ and\ \citenamefont
  {Gisin}}]{MasanesPRA2006}%
  \BibitemOpen
  \bibfield  {author} {\bibinfo {author} {\bibfnamefont {L.}~\bibnamefont
  {Masanes}}, \bibinfo {author} {\bibfnamefont {A.}~\bibnamefont {Ac\'in}}, \
  and\ \bibinfo {author} {\bibfnamefont {N.}~\bibnamefont {Gisin}},\ }\href
  {\doibase 10.1103/PhysRevA.73.012112} {\bibfield  {journal} {\bibinfo
  {journal} {Phys. Rev. A}\ }\textbf {\bibinfo {volume} {73}},\ \bibinfo
  {pages} {012112} (\bibinfo {year} {2006})}\BibitemShut {NoStop}%
\bibitem [{\citenamefont {Collins}\ and\ \citenamefont
  {Gisin}(2004)}]{Collins04}%
  \BibitemOpen
  \bibfield  {author} {\bibinfo {author} {\bibfnamefont {D.}~\bibnamefont
  {Collins}}\ and\ \bibinfo {author} {\bibfnamefont {N.}~\bibnamefont
  {Gisin}},\ }\href {https://doi.org/10.1088/0305-4470/37/5/021} {\bibfield
  {journal} {\bibinfo  {journal} {J. Phys. A: Math. Gen.}\ }\textbf {\bibinfo
  {volume} {37}},\ \bibinfo {pages} {1775} (\bibinfo {year}
  {2004})}\BibitemShut {NoStop}%
\bibitem [{\citenamefont {Scarani}\ and\ \citenamefont {Gisin}(2001)}]{SG01}%
  \BibitemOpen
  \bibfield  {author} {\bibinfo {author} {\bibfnamefont {V.}~\bibnamefont
  {Scarani}}\ and\ \bibinfo {author} {\bibfnamefont {N.}~\bibnamefont
  {Gisin}},\ }\href {\doibase 10.1103/PhysRevLett.87.117901} {\bibfield
  {journal} {\bibinfo  {journal} {Phys. Rev. Lett.}\ }\textbf {\bibinfo
  {volume} {87}},\ \bibinfo {pages} {117901} (\bibinfo {year}
  {2001})}\BibitemShut {NoStop}%
\bibitem [{\citenamefont {Toner}\ and\ \citenamefont
  {Verstraete}(2006)}]{TV2006}%
  \BibitemOpen
  \bibfield  {author} {\bibinfo {author} {\bibfnamefont {B.}~\bibnamefont
  {Toner}}\ and\ \bibinfo {author} {\bibfnamefont {F.}~\bibnamefont
  {Verstraete}},\ }\href {\doibase
  https://doi.org/10.48550/arXiv.quant-ph/0611001} {\enquote {\bibinfo {title}
  {Monogamy of {Bell} correlations and {Tsirelson's} bound},}\ }\bibinfo
  {howpublished} {arXiv preprint quant-ph/0611001} (\bibinfo {year}
  {2006})\BibitemShut {NoStop}%
\bibitem [{\citenamefont {Toner}(2009)}]{Toner:2008aa}%
  \BibitemOpen
  \bibfield  {author} {\bibinfo {author} {\bibfnamefont {B.}~\bibnamefont
  {Toner}},\ }\href {\doibase 10.1098/rspa.2008.0149} {\bibfield  {journal}
  {\bibinfo  {journal} {Proc. R. Soc. A}\ }\textbf {\bibinfo {volume} {465}},\
  \bibinfo {pages} {59} (\bibinfo {year} {2009})}\BibitemShut {NoStop}%
\bibitem [{\citenamefont {Paw\l{}owski}\ and\ \citenamefont
  {Brukner}(2009)}]{PB09}%
  \BibitemOpen
  \bibfield  {author} {\bibinfo {author} {\bibfnamefont {M.}~\bibnamefont
  {Paw\l{}owski}}\ and\ \bibinfo {author} {\bibfnamefont {{\v C}.}~\bibnamefont
  {Brukner}},\ }\href {\doibase 10.1103/PhysRevLett.102.030403} {\bibfield
  {journal} {\bibinfo  {journal} {Phys. Rev. Lett.}\ }\textbf {\bibinfo
  {volume} {102}},\ \bibinfo {pages} {030403} (\bibinfo {year}
  {2009})}\BibitemShut {NoStop}%
\bibitem [{\citenamefont {Kurzy\'{n}ski}\ \emph {et~al.}(2011)\citenamefont
  {Kurzy\'{n}ski}, \citenamefont {Paterek}, \citenamefont {Ramanathan},
  \citenamefont {Laskowski},\ and\ \citenamefont {Kaszlikowski}}]{KPR+11}%
  \BibitemOpen
  \bibfield  {author} {\bibinfo {author} {\bibfnamefont {P.}~\bibnamefont
  {Kurzy\'{n}ski}}, \bibinfo {author} {\bibfnamefont {T.}~\bibnamefont
  {Paterek}}, \bibinfo {author} {\bibfnamefont {R.}~\bibnamefont {Ramanathan}},
  \bibinfo {author} {\bibfnamefont {W.}~\bibnamefont {Laskowski}}, \ and\
  \bibinfo {author} {\bibfnamefont {D.}~\bibnamefont {Kaszlikowski}},\ }\href
  {\doibase 10.1103/PhysRevLett.106.180402} {\bibfield  {journal} {\bibinfo
  {journal} {Phys. Rev. Lett.}\ }\textbf {\bibinfo {volume} {106}},\ \bibinfo
  {pages} {180402} (\bibinfo {year} {2011})}\BibitemShut {NoStop}%
\bibitem [{\citenamefont {Ramanathan}\ and\ \citenamefont
  {Horodecki}(2014)}]{RH14}%
  \BibitemOpen
  \bibfield  {author} {\bibinfo {author} {\bibfnamefont {R.}~\bibnamefont
  {Ramanathan}}\ and\ \bibinfo {author} {\bibfnamefont {P.}~\bibnamefont
  {Horodecki}},\ }\href {\doibase 10.1103/PhysRevLett.113.210403} {\bibfield
  {journal} {\bibinfo  {journal} {Phys. Rev. Lett.}\ }\textbf {\bibinfo
  {volume} {113}},\ \bibinfo {pages} {210403} (\bibinfo {year}
  {2014})}\BibitemShut {NoStop}%
\bibitem [{\citenamefont {Yang}\ \emph {et~al.}(2024)\citenamefont {Yang},
  \citenamefont {Liu}, \citenamefont {Zheng}, \citenamefont {Fei},\ and\
  \citenamefont {Luo}}]{YLZ+24}%
  \BibitemOpen
  \bibfield  {author} {\bibinfo {author} {\bibfnamefont {Y.-H.}\ \bibnamefont
  {Yang}}, \bibinfo {author} {\bibfnamefont {X.-Z.}\ \bibnamefont {Liu}},
  \bibinfo {author} {\bibfnamefont {X.-Z.}\ \bibnamefont {Zheng}}, \bibinfo
  {author} {\bibfnamefont {S.-M.}\ \bibnamefont {Fei}}, \ and\ \bibinfo
  {author} {\bibfnamefont {M.-X.}\ \bibnamefont {Luo}},\ }\href {\doibase
  10.1016/j.xcrp.2024.101840} {\bibfield  {journal} {\bibinfo  {journal} {Cell
  Rep. Phys. Sci.}\ }\textbf {\bibinfo {volume} {5}},\ \bibinfo {pages}
  {101840} (\bibinfo {year} {2024})}\BibitemShut {NoStop}%
\bibitem [{\citenamefont {Cui}\ \emph {et~al.}(2024)\citenamefont {Cui},
  \citenamefont {Mehta},\ and\ \citenamefont {Rochette}}]{CMR24}%
  \BibitemOpen
  \bibfield  {author} {\bibinfo {author} {\bibfnamefont {D.}~\bibnamefont
  {Cui}}, \bibinfo {author} {\bibfnamefont {A.}~\bibnamefont {Mehta}}, \ and\
  \bibinfo {author} {\bibfnamefont {D.}~\bibnamefont {Rochette}},\ }\href@noop
  {} {\enquote {\bibinfo {title} {Monogamy of nonlocal games},}\ } (\bibinfo
  {year} {2024}),\ \Eprint {http://arxiv.org/abs/2405.20286v3}
  {arXiv:2405.20286v3 [quant-ph]} \BibitemShut {NoStop}%
\bibitem [{\citenamefont {Tabia}\ \emph {et~al.}(2022)\citenamefont {Tabia},
  \citenamefont {Chen}, \citenamefont {Hsieh}, \citenamefont {Yin},\ and\
  \citenamefont {Liang}}]{Tabia2022}%
  \BibitemOpen
  \bibfield  {author} {\bibinfo {author} {\bibfnamefont {G.~N.~M.}\
  \bibnamefont {Tabia}}, \bibinfo {author} {\bibfnamefont {K.-S.}\ \bibnamefont
  {Chen}}, \bibinfo {author} {\bibfnamefont {C.-Y.}\ \bibnamefont {Hsieh}},
  \bibinfo {author} {\bibfnamefont {Y.-C.}\ \bibnamefont {Yin}}, \ and\
  \bibinfo {author} {\bibfnamefont {Y.-C.}\ \bibnamefont {Liang}},\ }\href
  {\doibase 10.1038/s41534-022-00616-1} {\bibfield  {journal} {\bibinfo
  {journal} {npj Quantum Inf.}\ }\textbf {\bibinfo {volume} {8}},\ \bibinfo
  {pages} {98} (\bibinfo {year} {2022})}\BibitemShut {NoStop}%
\bibitem [{\citenamefont {Coretti}\ \emph {et~al.}(2011)\citenamefont
  {Coretti}, \citenamefont {H\"anggi},\ and\ \citenamefont
  {Wolf}}]{Coretti2011}%
  \BibitemOpen
  \bibfield  {author} {\bibinfo {author} {\bibfnamefont {S.}~\bibnamefont
  {Coretti}}, \bibinfo {author} {\bibfnamefont {E.}~\bibnamefont {H\"anggi}}, \
  and\ \bibinfo {author} {\bibfnamefont {S.}~\bibnamefont {Wolf}},\ }\href
  {\doibase 10.1103/PhysRevLett.107.100402} {\bibfield  {journal} {\bibinfo
  {journal} {Phys. Rev. Lett.}\ }\textbf {\bibinfo {volume} {107}},\ \bibinfo
  {pages} {100402} (\bibinfo {year} {2011})}\BibitemShut {NoStop}%
\bibitem [{\citenamefont {Scarani}\ and\ \citenamefont
  {Gisin}(2002)}]{Scarani:2002aa}%
  \BibitemOpen
  \bibfield  {author} {\bibinfo {author} {\bibfnamefont {V.}~\bibnamefont
  {Scarani}}\ and\ \bibinfo {author} {\bibfnamefont {N.}~\bibnamefont
  {Gisin}},\ }\href {\doibase https://doi.org/10.1016/S0375-9601(02)00174-3}
  {\bibfield  {journal} {\bibinfo  {journal} {Phys. Lett. A}\ }\textbf
  {\bibinfo {volume} {295}},\ \bibinfo {pages} {167} (\bibinfo {year}
  {2002})}\BibitemShut {NoStop}%
\bibitem [{\citenamefont {Scarani}\ and\ \citenamefont
  {Gisin}(2005)}]{scarani_gisin_2005}%
  \BibitemOpen
  \bibfield  {author} {\bibinfo {author} {\bibfnamefont {V.}~\bibnamefont
  {Scarani}}\ and\ \bibinfo {author} {\bibfnamefont {N.}~\bibnamefont
  {Gisin}},\ }\href {\doibase 10.1590/S0103-97332005000200018} {\bibfield
  {journal} {\bibinfo  {journal} {Braz. J. Phys.}\ }\textbf {\bibinfo {volume}
  {35}},\ \bibinfo {pages} {328} (\bibinfo {year} {2005})}\BibitemShut
  {NoStop}%
\bibitem [{\citenamefont {Bancal}\ \emph {et~al.}(2012)\citenamefont {Bancal},
  \citenamefont {Pironio}, \citenamefont {Ac{\'\i}n}, \citenamefont {Liang},
  \citenamefont {Scarani},\ and\ \citenamefont {Gisin}}]{Bancal2012}%
  \BibitemOpen
  \bibfield  {author} {\bibinfo {author} {\bibfnamefont {J.-D.}\ \bibnamefont
  {Bancal}}, \bibinfo {author} {\bibfnamefont {S.}~\bibnamefont {Pironio}},
  \bibinfo {author} {\bibfnamefont {A.}~\bibnamefont {Ac{\'\i}n}}, \bibinfo
  {author} {\bibfnamefont {Y.-C.}\ \bibnamefont {Liang}}, \bibinfo {author}
  {\bibfnamefont {V.}~\bibnamefont {Scarani}}, \ and\ \bibinfo {author}
  {\bibfnamefont {N.}~\bibnamefont {Gisin}},\ }\href {\doibase
  10.1038/nphys2460} {\bibfield  {journal} {\bibinfo  {journal} {Nat. Phys.}\
  }\textbf {\bibinfo {volume} {8}},\ \bibinfo {pages} {867} (\bibinfo {year}
  {2012})}\BibitemShut {NoStop}%
\bibitem [{\citenamefont {Barnea}\ \emph {et~al.}(2013)\citenamefont {Barnea},
  \citenamefont {Bancal}, \citenamefont {Liang},\ and\ \citenamefont
  {Gisin}}]{Barnea2013}%
  \BibitemOpen
  \bibfield  {author} {\bibinfo {author} {\bibfnamefont {T.~J.}\ \bibnamefont
  {Barnea}}, \bibinfo {author} {\bibfnamefont {J.-D.}\ \bibnamefont {Bancal}},
  \bibinfo {author} {\bibfnamefont {Y.-C.}\ \bibnamefont {Liang}}, \ and\
  \bibinfo {author} {\bibfnamefont {N.}~\bibnamefont {Gisin}},\ }\href
  {\doibase 10.1103/PhysRevA.88.022123} {\bibfield  {journal} {\bibinfo
  {journal} {Phys. Rev. A}\ }\textbf {\bibinfo {volume} {88}},\ \bibinfo
  {pages} {022123} (\bibinfo {year} {2013})}\BibitemShut {NoStop}%
\bibitem [{\citenamefont {Chiribella}\ and\ \citenamefont
  {Spekkens}(2016)}]{Foil2016}%
  \BibitemOpen
  \bibinfo {editor} {\bibfnamefont {G.}~\bibnamefont {Chiribella}}\ and\
  \bibinfo {editor} {\bibfnamefont {R.~W.}\ \bibnamefont {Spekkens}},\ eds.,\
  \href@noop {} {\emph {\bibinfo {title} {Quantum Theory: Informational
  Foundations and Foils}}}\ (\bibinfo  {publisher} {Springer},\ \bibinfo
  {address} {Dordrecht},\ \bibinfo {year} {2016})\BibitemShut {NoStop}%
\bibitem [{\citenamefont {Brunner}\ and\ \citenamefont
  {V\'ertesi}(2012)}]{BV2012}%
  \BibitemOpen
  \bibfield  {author} {\bibinfo {author} {\bibfnamefont {N.}~\bibnamefont
  {Brunner}}\ and\ \bibinfo {author} {\bibfnamefont {T.}~\bibnamefont
  {V\'ertesi}},\ }\href {\doibase 10.1103/PhysRevA.86.042113} {\bibfield
  {journal} {\bibinfo  {journal} {Phys. Rev. A}\ }\textbf {\bibinfo {volume}
  {86}},\ \bibinfo {pages} {042113} (\bibinfo {year} {2012})}\BibitemShut
  {NoStop}%
\bibitem [{\citenamefont {Khot}\ and\ \citenamefont
  {Vishnoi}(2005)}]{KhotVishnoi2005}%
  \BibitemOpen
  \bibfield  {author} {\bibinfo {author} {\bibfnamefont {S.}~\bibnamefont
  {Khot}}\ and\ \bibinfo {author} {\bibfnamefont {N.}~\bibnamefont {Vishnoi}},\
  }in\ \href {\doibase 10.1109/SFCS.2005.74} {\emph {\bibinfo {booktitle} {46th
  Annual IEEE Symposium on Foundations of Computer Science (FOCS'05)}}}\
  (\bibinfo {year} {2005})\ pp.\ \bibinfo {pages} {53--62}\BibitemShut
  {NoStop}%
\bibitem [{\citenamefont {Palazuelos}(2012)}]{Palazuelos2012}%
  \BibitemOpen
  \bibfield  {author} {\bibinfo {author} {\bibfnamefont {C.}~\bibnamefont
  {Palazuelos}},\ }\href {\doibase 10.1103/PhysRevLett.109.190401} {\bibfield
  {journal} {\bibinfo  {journal} {Phys. Rev. Lett.}\ }\textbf {\bibinfo
  {volume} {109}},\ \bibinfo {pages} {190401} (\bibinfo {year}
  {2012})}\BibitemShut {NoStop}%
\bibitem [{\citenamefont {Cavalcanti}\ \emph {et~al.}(2013)\citenamefont
  {Cavalcanti}, \citenamefont {Ac\'{\i}n}, \citenamefont {Brunner},\ and\
  \citenamefont {V\'ertesi}}]{Cavalcanti2013}%
  \BibitemOpen
  \bibfield  {author} {\bibinfo {author} {\bibfnamefont {D.}~\bibnamefont
  {Cavalcanti}}, \bibinfo {author} {\bibfnamefont {A.}~\bibnamefont
  {Ac\'{\i}n}}, \bibinfo {author} {\bibfnamefont {N.}~\bibnamefont {Brunner}},
  \ and\ \bibinfo {author} {\bibfnamefont {T.}~\bibnamefont {V\'ertesi}},\
  }\href {\doibase 10.1103/PhysRevA.87.042104} {\bibfield  {journal} {\bibinfo
  {journal} {Phys. Rev. A}\ }\textbf {\bibinfo {volume} {87}},\ \bibinfo
  {pages} {042104} (\bibinfo {year} {2013})}\BibitemShut {NoStop}%
\bibitem [{\citenamefont {Nielsen}\ and\ \citenamefont
  {Chuang}(2000)}]{QCI-book}%
  \BibitemOpen
  \bibfield  {author} {\bibinfo {author} {\bibfnamefont {M.~A.}\ \bibnamefont
  {Nielsen}}\ and\ \bibinfo {author} {\bibfnamefont {I.~L.}\ \bibnamefont
  {Chuang}},\ }\href@noop {} {\emph {\bibinfo {title} {{Quantum Computation and
  Quantum Information}}}}\ (\bibinfo  {publisher} {Cambridge University
  Press},\ \bibinfo {address} {Cambridge},\ \bibinfo {year} {2000})\BibitemShut
  {NoStop}%
\bibitem [{\citenamefont {Cleve}\ \emph {et~al.}(2004)\citenamefont {Cleve},
  \citenamefont {Hoyer}, \citenamefont {Toner},\ and\ \citenamefont
  {Watrous}}]{Cleve:IEEE:2004}%
  \BibitemOpen
  \bibfield  {author} {\bibinfo {author} {\bibfnamefont {R.}~\bibnamefont
  {Cleve}}, \bibinfo {author} {\bibfnamefont {P.}~\bibnamefont {Hoyer}},
  \bibinfo {author} {\bibfnamefont {B.}~\bibnamefont {Toner}}, \ and\ \bibinfo
  {author} {\bibfnamefont {J.}~\bibnamefont {Watrous}},\ }in\ \href {\doibase
  10.1109/CCC.2004.1313847} {\emph {\bibinfo {booktitle} {Proceedings. 19th
  IEEE Annual Conference on Computational Complexity, 2004.}}}\ (\bibinfo
  {year} {2004})\ pp.\ \bibinfo {pages} {236--249}\BibitemShut {NoStop}%
\bibitem [{\citenamefont {Clauser}\ \emph {et~al.}(1969)\citenamefont
  {Clauser}, \citenamefont {Horne}, \citenamefont {Shimony},\ and\
  \citenamefont {Holt}}]{CHSH}%
  \BibitemOpen
  \bibfield  {author} {\bibinfo {author} {\bibfnamefont {J.~F.}\ \bibnamefont
  {Clauser}}, \bibinfo {author} {\bibfnamefont {M.~A.}\ \bibnamefont {Horne}},
  \bibinfo {author} {\bibfnamefont {A.}~\bibnamefont {Shimony}}, \ and\
  \bibinfo {author} {\bibfnamefont {R.~A.}\ \bibnamefont {Holt}},\ }\href
  {\doibase 10.1103/PhysRevLett.23.880} {\bibfield  {journal} {\bibinfo
  {journal} {Phys. Rev. Lett.}\ }\textbf {\bibinfo {volume} {23}},\ \bibinfo
  {pages} {880} (\bibinfo {year} {1969})}\BibitemShut {NoStop}%
\bibitem [{\citenamefont {Brassard}\ \emph {et~al.}(2006)\citenamefont
  {Brassard}, \citenamefont {Buhrman}, \citenamefont {Linden}, \citenamefont
  {M\'ethot}, \citenamefont {Tapp},\ and\ \citenamefont {Unger}}]{BBL+06}%
  \BibitemOpen
  \bibfield  {author} {\bibinfo {author} {\bibfnamefont {G.}~\bibnamefont
  {Brassard}}, \bibinfo {author} {\bibfnamefont {H.}~\bibnamefont {Buhrman}},
  \bibinfo {author} {\bibfnamefont {N.}~\bibnamefont {Linden}}, \bibinfo
  {author} {\bibfnamefont {A.~A.}\ \bibnamefont {M\'ethot}}, \bibinfo {author}
  {\bibfnamefont {A.}~\bibnamefont {Tapp}}, \ and\ \bibinfo {author}
  {\bibfnamefont {F.}~\bibnamefont {Unger}},\ }\href {\doibase
  10.1103/PhysRevLett.96.250401} {\bibfield  {journal} {\bibinfo  {journal}
  {Phys. Rev. Lett.}\ }\textbf {\bibinfo {volume} {96}},\ \bibinfo {pages}
  {250401} (\bibinfo {year} {2006})}\BibitemShut {NoStop}%
\bibitem [{\citenamefont {Paw{\l}owski}\ \emph {et~al.}(2009)\citenamefont
  {Paw{\l}owski}, \citenamefont {Paterek}, \citenamefont {Kaszlikowski},
  \citenamefont {Scarani}, \citenamefont {Winter},\ and\ \citenamefont
  {{\.Z}ukowski}}]{Pawowski:2009aa}%
  \BibitemOpen
  \bibfield  {author} {\bibinfo {author} {\bibfnamefont {M.}~\bibnamefont
  {Paw{\l}owski}}, \bibinfo {author} {\bibfnamefont {T.}~\bibnamefont
  {Paterek}}, \bibinfo {author} {\bibfnamefont {D.}~\bibnamefont
  {Kaszlikowski}}, \bibinfo {author} {\bibfnamefont {V.}~\bibnamefont
  {Scarani}}, \bibinfo {author} {\bibfnamefont {A.}~\bibnamefont {Winter}}, \
  and\ \bibinfo {author} {\bibfnamefont {M.}~\bibnamefont {{\.Z}ukowski}},\
  }\href {\doibase 10.1038/nature08400} {\bibfield  {journal} {\bibinfo
  {journal} {Nature}\ }\textbf {\bibinfo {volume} {461}},\ \bibinfo {pages}
  {1101} (\bibinfo {year} {2009})}\BibitemShut {NoStop}%
\bibitem [{\citenamefont {Navascu{\'e}s}\ and\ \citenamefont
  {Wunderlich}(2009)}]{Navascues:2009aa}%
  \BibitemOpen
  \bibfield  {author} {\bibinfo {author} {\bibfnamefont {M.}~\bibnamefont
  {Navascu{\'e}s}}\ and\ \bibinfo {author} {\bibfnamefont {H.}~\bibnamefont
  {Wunderlich}},\ }\href {\doibase 10.1098/rspa.2009.0453} {\bibfield
  {journal} {\bibinfo  {journal} {Proc. R. Soc. A: Math. Phys. Eng. Sci.}\
  }\textbf {\bibinfo {volume} {466}},\ \bibinfo {pages} {881} (\bibinfo {year}
  {2009})}\BibitemShut {NoStop}%
\bibitem [{\citenamefont {Sainz}\ \emph {et~al.}(2018)\citenamefont {Sainz},
  \citenamefont {Guryanova}, \citenamefont {Ac\'{\i}n},\ and\ \citenamefont
  {Navascu\'es}}]{SGA+18}%
  \BibitemOpen
  \bibfield  {author} {\bibinfo {author} {\bibfnamefont {A.~B.}\ \bibnamefont
  {Sainz}}, \bibinfo {author} {\bibfnamefont {Y.}~\bibnamefont {Guryanova}},
  \bibinfo {author} {\bibfnamefont {A.}~\bibnamefont {Ac\'{\i}n}}, \ and\
  \bibinfo {author} {\bibfnamefont {M.}~\bibnamefont {Navascu\'es}},\ }\href
  {\doibase 10.1103/PhysRevLett.120.200402} {\bibfield  {journal} {\bibinfo
  {journal} {Phys. Rev. Lett.}\ }\textbf {\bibinfo {volume} {120}},\ \bibinfo
  {pages} {200402} (\bibinfo {year} {2018})}\BibitemShut {NoStop}%
\bibitem [{\citenamefont {Werner}(1989)}]{Werner:PRA:1989}%
  \BibitemOpen
  \bibfield  {author} {\bibinfo {author} {\bibfnamefont {R.~F.}\ \bibnamefont
  {Werner}},\ }\href {\doibase 10.1103/PhysRevA.40.4277} {\bibfield  {journal}
  {\bibinfo  {journal} {Phys. Rev. A}\ }\textbf {\bibinfo {volume} {40}},\
  \bibinfo {pages} {4277} (\bibinfo {year} {1989})}\BibitemShut {NoStop}%
\bibitem [{\citenamefont {Terhal}\ \emph {et~al.}(2003)\citenamefont {Terhal},
  \citenamefont {Doherty},\ and\ \citenamefont {Schwab}}]{Terhal03}%
  \BibitemOpen
  \bibfield  {author} {\bibinfo {author} {\bibfnamefont {B.~M.}\ \bibnamefont
  {Terhal}}, \bibinfo {author} {\bibfnamefont {A.~C.}\ \bibnamefont {Doherty}},
  \ and\ \bibinfo {author} {\bibfnamefont {D.}~\bibnamefont {Schwab}},\ }\href
  {\doibase 10.1103/PhysRevLett.90.157903} {\bibfield  {journal} {\bibinfo
  {journal} {Phys. Rev. Lett.}\ }\textbf {\bibinfo {volume} {90}},\ \bibinfo
  {pages} {157903} (\bibinfo {year} {2003})}\BibitemShut {NoStop}%
\bibitem [{\citenamefont {Liang}\ and\ \citenamefont {Doherty}(2007)}]{LD07}%
  \BibitemOpen
  \bibfield  {author} {\bibinfo {author} {\bibfnamefont {Y.-C.}\ \bibnamefont
  {Liang}}\ and\ \bibinfo {author} {\bibfnamefont {A.~C.}\ \bibnamefont
  {Doherty}},\ }\href {\doibase 10.1103/PhysRevA.75.042103} {\bibfield
  {journal} {\bibinfo  {journal} {Phys. Rev. A}\ }\textbf {\bibinfo {volume}
  {75}},\ \bibinfo {pages} {042103} (\bibinfo {year} {2007})}\BibitemShut
  {NoStop}%
\bibitem [{\citenamefont {Hirsch}\ \emph {et~al.}(2016)\citenamefont {Hirsch},
  \citenamefont {Quintino}, \citenamefont {V\'ertesi}, \citenamefont {Pusey},\
  and\ \citenamefont {Brunner}}]{HirschPRL16}%
  \BibitemOpen
  \bibfield  {author} {\bibinfo {author} {\bibfnamefont {F.}~\bibnamefont
  {Hirsch}}, \bibinfo {author} {\bibfnamefont {M.~T.}\ \bibnamefont
  {Quintino}}, \bibinfo {author} {\bibfnamefont {T.}~\bibnamefont {V\'ertesi}},
  \bibinfo {author} {\bibfnamefont {M.~F.}\ \bibnamefont {Pusey}}, \ and\
  \bibinfo {author} {\bibfnamefont {N.}~\bibnamefont {Brunner}},\ }\href
  {\doibase 10.1103/PhysRevLett.117.190402} {\bibfield  {journal} {\bibinfo
  {journal} {Phys. Rev. Lett.}\ }\textbf {\bibinfo {volume} {117}},\ \bibinfo
  {pages} {190402} (\bibinfo {year} {2016})}\BibitemShut {NoStop}%
\bibitem [{\citenamefont {Cavalcanti}\ \emph {et~al.}(2016)\citenamefont
  {Cavalcanti}, \citenamefont {Guerini}, \citenamefont {Rabelo},\ and\
  \citenamefont {Skrzypczyk}}]{CavalcantiPRL16}%
  \BibitemOpen
  \bibfield  {author} {\bibinfo {author} {\bibfnamefont {D.}~\bibnamefont
  {Cavalcanti}}, \bibinfo {author} {\bibfnamefont {L.}~\bibnamefont {Guerini}},
  \bibinfo {author} {\bibfnamefont {R.}~\bibnamefont {Rabelo}}, \ and\ \bibinfo
  {author} {\bibfnamefont {P.}~\bibnamefont {Skrzypczyk}},\ }\href {\doibase
  10.1103/PhysRevLett.117.190401} {\bibfield  {journal} {\bibinfo  {journal}
  {Phys. Rev. Lett.}\ }\textbf {\bibinfo {volume} {117}},\ \bibinfo {pages}
  {190401} (\bibinfo {year} {2016})}\BibitemShut {NoStop}%
\bibitem [{\citenamefont {Hsieh}\ \emph {et~al.}(2016)\citenamefont {Hsieh},
  \citenamefont {Liang},\ and\ \citenamefont {Lee}}]{Hsieh2016}%
  \BibitemOpen
  \bibfield  {author} {\bibinfo {author} {\bibfnamefont {C.-Y.}\ \bibnamefont
  {Hsieh}}, \bibinfo {author} {\bibfnamefont {Y.-C.}\ \bibnamefont {Liang}}, \
  and\ \bibinfo {author} {\bibfnamefont {R.-K.}\ \bibnamefont {Lee}},\ }\href
  {\doibase 10.1103/PhysRevA.94.062120} {\bibfield  {journal} {\bibinfo
  {journal} {Phys. Rev. A}\ }\textbf {\bibinfo {volume} {94}},\ \bibinfo
  {pages} {062120} (\bibinfo {year} {2016})}\BibitemShut {NoStop}%
\bibitem [{\citenamefont {Horodecki}\ \emph {et~al.}(1995)\citenamefont
  {Horodecki}, \citenamefont {Horodecki},\ and\ \citenamefont
  {Horodecki}}]{HHH95}%
  \BibitemOpen
  \bibfield  {author} {\bibinfo {author} {\bibfnamefont {R.}~\bibnamefont
  {Horodecki}}, \bibinfo {author} {\bibfnamefont {P.}~\bibnamefont
  {Horodecki}}, \ and\ \bibinfo {author} {\bibfnamefont {M.}~\bibnamefont
  {Horodecki}},\ }\href {\doibase https://doi.org/10.1016/0375-9601(95)00214-N}
  {\bibfield  {journal} {\bibinfo  {journal} {Phys. Lett. A}\ }\textbf
  {\bibinfo {volume} {200}},\ \bibinfo {pages} {340} (\bibinfo {year}
  {1995})}\BibitemShut {NoStop}%
\bibitem [{\citenamefont {D\"ur}\ \emph {et~al.}(2000)\citenamefont {D\"ur},
  \citenamefont {Vidal},\ and\ \citenamefont {Cirac}}]{Dur2000}%
  \BibitemOpen
  \bibfield  {author} {\bibinfo {author} {\bibfnamefont {W.}~\bibnamefont
  {D\"ur}}, \bibinfo {author} {\bibfnamefont {G.}~\bibnamefont {Vidal}}, \ and\
  \bibinfo {author} {\bibfnamefont {J.~I.}\ \bibnamefont {Cirac}},\ }\href
  {\doibase 10.1103/PhysRevA.62.062314} {\bibfield  {journal} {\bibinfo
  {journal} {Phys. Rev. A}\ }\textbf {\bibinfo {volume} {62}},\ \bibinfo
  {pages} {062314} (\bibinfo {year} {2000})}\BibitemShut {NoStop}%
\bibitem [{\citenamefont {Haddara}()}]{PrivateMarwan}%
  \BibitemOpen
  \bibfield  {author} {\bibinfo {author} {\bibfnamefont {M.}~\bibnamefont
  {Haddara}},\ }\href@noop {} {}\bibinfo {howpublished} {(private
  communication)}\BibitemShut {NoStop}%
\bibitem [{\citenamefont {T\'oth}\ \emph {et~al.}(2007)\citenamefont {T\'oth},
  \citenamefont {Knapp}, \citenamefont {G\"uhne},\ and\ \citenamefont
  {Briegel}}]{Toth2007}%
  \BibitemOpen
  \bibfield  {author} {\bibinfo {author} {\bibfnamefont {G.}~\bibnamefont
  {T\'oth}}, \bibinfo {author} {\bibfnamefont {C.}~\bibnamefont {Knapp}},
  \bibinfo {author} {\bibfnamefont {O.}~\bibnamefont {G\"uhne}}, \ and\
  \bibinfo {author} {\bibfnamefont {H.~J.}\ \bibnamefont {Briegel}},\ }\href
  {\doibase 10.1103/PhysRevLett.99.250405} {\bibfield  {journal} {\bibinfo
  {journal} {Phys. Rev. Lett.}\ }\textbf {\bibinfo {volume} {99}},\ \bibinfo
  {pages} {250405} (\bibinfo {year} {2007})}\BibitemShut {NoStop}%
\bibitem [{\citenamefont {W\"urflinger}\ \emph {et~al.}(2012)\citenamefont
  {W\"urflinger}, \citenamefont {Bancal}, \citenamefont {Ac\'{\i}n},
  \citenamefont {Gisin},\ and\ \citenamefont {V\'ertesi}}]{WBA+12}%
  \BibitemOpen
  \bibfield  {author} {\bibinfo {author} {\bibfnamefont {L.~E.}\ \bibnamefont
  {W\"urflinger}}, \bibinfo {author} {\bibfnamefont {J.-D.}\ \bibnamefont
  {Bancal}}, \bibinfo {author} {\bibfnamefont {A.}~\bibnamefont {Ac\'{\i}n}},
  \bibinfo {author} {\bibfnamefont {N.}~\bibnamefont {Gisin}}, \ and\ \bibinfo
  {author} {\bibfnamefont {T.}~\bibnamefont {V\'ertesi}},\ }\href {\doibase
  10.1103/PhysRevA.86.032117} {\bibfield  {journal} {\bibinfo  {journal} {Phys.
  Rev. A}\ }\textbf {\bibinfo {volume} {86}},\ \bibinfo {pages} {032117}
  (\bibinfo {year} {2012})}\BibitemShut {NoStop}%
\bibitem [{\citenamefont {Liang}\ \emph {et~al.}(2014)\citenamefont {Liang},
  \citenamefont {Curchod}, \citenamefont {Bowles},\ and\ \citenamefont
  {Gisin}}]{LCBG14}%
  \BibitemOpen
  \bibfield  {author} {\bibinfo {author} {\bibfnamefont {Y.-C.}\ \bibnamefont
  {Liang}}, \bibinfo {author} {\bibfnamefont {F.~J.}\ \bibnamefont {Curchod}},
  \bibinfo {author} {\bibfnamefont {J.}~\bibnamefont {Bowles}}, \ and\ \bibinfo
  {author} {\bibfnamefont {N.}~\bibnamefont {Gisin}},\ }\href {\doibase
  10.1103/PhysRevLett.113.130401} {\bibfield  {journal} {\bibinfo  {journal}
  {Phys. Rev. Lett.}\ }\textbf {\bibinfo {volume} {113}},\ \bibinfo {pages}
  {130401} (\bibinfo {year} {2014})}\BibitemShut {NoStop}%
\bibitem [{\citenamefont {Navascu{\'{e}}s}\ \emph {et~al.}(2021)\citenamefont
  {Navascu{\'{e}}s}, \citenamefont {Baccari},\ and\ \citenamefont
  {Ac{\'{i}}n}}]{Navascues2021}%
  \BibitemOpen
  \bibfield  {author} {\bibinfo {author} {\bibfnamefont {M.}~\bibnamefont
  {Navascu{\'{e}}s}}, \bibinfo {author} {\bibfnamefont {F.}~\bibnamefont
  {Baccari}}, \ and\ \bibinfo {author} {\bibfnamefont {A.}~\bibnamefont
  {Ac{\'{i}}n}},\ }\href {\doibase 10.22331/q-2021-11-25-589} {\bibfield
  {journal} {\bibinfo  {journal} {{Quantum}}\ }\textbf {\bibinfo {volume}
  {5}},\ \bibinfo {pages} {589} (\bibinfo {year} {2021})}\BibitemShut {NoStop}%
\bibitem [{\citenamefont {Bancal}\ \emph {et~al.}(2015)\citenamefont {Bancal},
  \citenamefont {Navascu\'es}, \citenamefont {Scarani}, \citenamefont
  {V\'ertesi},\ and\ \citenamefont {Yang}}]{BNS+15}%
  \BibitemOpen
  \bibfield  {author} {\bibinfo {author} {\bibfnamefont {J.-D.}\ \bibnamefont
  {Bancal}}, \bibinfo {author} {\bibfnamefont {M.}~\bibnamefont {Navascu\'es}},
  \bibinfo {author} {\bibfnamefont {V.}~\bibnamefont {Scarani}}, \bibinfo
  {author} {\bibfnamefont {T.}~\bibnamefont {V\'ertesi}}, \ and\ \bibinfo
  {author} {\bibfnamefont {T.~H.}\ \bibnamefont {Yang}},\ }\href {\doibase
  10.1103/PhysRevA.91.022115} {\bibfield  {journal} {\bibinfo  {journal} {Phys.
  Rev. A}\ }\textbf {\bibinfo {volume} {91}},\ \bibinfo {pages} {022115}
  (\bibinfo {year} {2015})}\BibitemShut {NoStop}%
\bibitem [{\citenamefont {Bender}\ and\ \citenamefont
  {Williamson}(2010)}]{BenderWilliamson2010}%
  \BibitemOpen
  \bibfield  {author} {\bibinfo {author} {\bibfnamefont {E.~A.}\ \bibnamefont
  {Bender}}\ and\ \bibinfo {author} {\bibfnamefont {S.~G.}\ \bibnamefont
  {Williamson}},\ }\enquote {\bibinfo {title} {{Lists, Decisions and
  Graphs}},}\ \ (\bibinfo  {publisher} {UC San Diego},\ \bibinfo {year}
  {2010})\ p.\ \bibinfo {pages} {171}\BibitemShut {NoStop}%
\bibitem [{\citenamefont {Parashar}\ and\ \citenamefont
  {Rana}(2009)}]{Parashar.PRA.2009}%
  \BibitemOpen
  \bibfield  {author} {\bibinfo {author} {\bibfnamefont {P.}~\bibnamefont
  {Parashar}}\ and\ \bibinfo {author} {\bibfnamefont {S.}~\bibnamefont
  {Rana}},\ }\href {\doibase 10.1103/PhysRevA.80.012319} {\bibfield  {journal}
  {\bibinfo  {journal} {Phys. Rev. A}\ }\textbf {\bibinfo {volume} {80}},\
  \bibinfo {pages} {012319} (\bibinfo {year} {2009})}\BibitemShut {NoStop}%
\bibitem [{\citenamefont {Wu}\ \emph {et~al.}(2014)\citenamefont {Wu},
  \citenamefont {Tian}, \citenamefont {Huang}, \citenamefont {Wen},
  \citenamefont {Qin},\ and\ \citenamefont {Gao}}]{Wu2014}%
  \BibitemOpen
  \bibfield  {author} {\bibinfo {author} {\bibfnamefont {X.}~\bibnamefont
  {Wu}}, \bibinfo {author} {\bibfnamefont {G.-J.}\ \bibnamefont {Tian}},
  \bibinfo {author} {\bibfnamefont {W.}~\bibnamefont {Huang}}, \bibinfo
  {author} {\bibfnamefont {Q.-Y.}\ \bibnamefont {Wen}}, \bibinfo {author}
  {\bibfnamefont {S.-J.}\ \bibnamefont {Qin}}, \ and\ \bibinfo {author}
  {\bibfnamefont {F.}~\bibnamefont {Gao}},\ }\href {\doibase
  10.1103/PhysRevA.90.012317} {\bibfield  {journal} {\bibinfo  {journal} {Phys.
  Rev. A}\ }\textbf {\bibinfo {volume} {90}},\ \bibinfo {pages} {012317}
  (\bibinfo {year} {2014})}\BibitemShut {NoStop}%
\bibitem [{\citenamefont {Shen}\ and\ \citenamefont {Chen}(2023)}]{SC23}%
  \BibitemOpen
  \bibfield  {author} {\bibinfo {author} {\bibfnamefont {Y.}~\bibnamefont
  {Shen}}\ and\ \bibinfo {author} {\bibfnamefont {L.}~\bibnamefont {Chen}},\
  }\href {\doibase 10.1103/PhysRevA.108.062418} {\bibfield  {journal} {\bibinfo
   {journal} {Phys. Rev. A}\ }\textbf {\bibinfo {volume} {108}},\ \bibinfo
  {pages} {062418} (\bibinfo {year} {2023})}\BibitemShut {NoStop}%
\bibitem [{\citenamefont {Horodecki}\ and\ \citenamefont
  {Horodecki}(1999)}]{HorodeckiMP1999}%
  \BibitemOpen
  \bibfield  {author} {\bibinfo {author} {\bibfnamefont {M.}~\bibnamefont
  {Horodecki}}\ and\ \bibinfo {author} {\bibfnamefont {P.}~\bibnamefont
  {Horodecki}},\ }\href {\doibase 10.1103/PhysRevA.59.4206} {\bibfield
  {journal} {\bibinfo  {journal} {Phys. Rev. A}\ }\textbf {\bibinfo {volume}
  {59}},\ \bibinfo {pages} {4206} (\bibinfo {year} {1999})}\BibitemShut
  {NoStop}%
\bibitem{Liu2024}%
M.-E. Liu, K.-S. Chen, C.-Y. Hsieh, G. N. M. Tabia, and Y.-C.
Liang, ``Large parts are generically entangled across all cuts,'' 
\Eprint {http://arxiv.org/abs/2505.20420v1}
  {arXiv:2505.20420v1 [quant-ph].}
\bibitem [{\citenamefont {Chen}\ \emph {et~al.}(2013)\citenamefont {Chen},
  \citenamefont {Dawkins}, \citenamefont {Ji}, \citenamefont {Johnston},
  \citenamefont {Kribs}, \citenamefont {Shultz},\ and\ \citenamefont
  {Zeng}}]{CDJ+13}%
  \BibitemOpen
  \bibfield  {author} {\bibinfo {author} {\bibfnamefont {J.}~\bibnamefont
  {Chen}}, \bibinfo {author} {\bibfnamefont {H.}~\bibnamefont {Dawkins}},
  \bibinfo {author} {\bibfnamefont {Z.}~\bibnamefont {Ji}}, \bibinfo {author}
  {\bibfnamefont {N.}~\bibnamefont {Johnston}}, \bibinfo {author}
  {\bibfnamefont {D.}~\bibnamefont {Kribs}}, \bibinfo {author} {\bibfnamefont
  {F.}~\bibnamefont {Shultz}}, \ and\ \bibinfo {author} {\bibfnamefont
  {B.}~\bibnamefont {Zeng}},\ }\href {\doibase 10.1103/PhysRevA.88.012109}
  {\bibfield  {journal} {\bibinfo  {journal} {Phys. Rev. A}\ }\textbf {\bibinfo
  {volume} {88}},\ \bibinfo {pages} {012109} (\bibinfo {year}
  {2013})}\BibitemShut {NoStop}%
\bibitem [{git()}]{githubCode}%
  \BibitemOpen
  \href@noop {} {}\bibinfo {howpublished}
  {\url{https://github.com/ycliangTW/NonlocalityTransitivity}}\BibitemShut
  {NoStop}%
\bibitem [{\citenamefont {Liang}\ \emph {et~al.}(2011)\citenamefont {Liang},
  \citenamefont {Spekkens},\ and\ \citenamefont {Wiseman}}]{Liang:PRep}%
  \BibitemOpen
  \bibfield  {author} {\bibinfo {author} {\bibfnamefont {Y.-C.}\ \bibnamefont
  {Liang}}, \bibinfo {author} {\bibfnamefont {R.~W.}\ \bibnamefont {Spekkens}},
  \ and\ \bibinfo {author} {\bibfnamefont {H.~M.}\ \bibnamefont {Wiseman}},\
  }\href {\doibase 10.1016/j.physrep.2011.05.001} {\bibfield  {journal}
  {\bibinfo  {journal} {Phys. Rep.}\ }\textbf {\bibinfo {volume} {506}},\
  \bibinfo {pages} {1 } (\bibinfo {year} {2011})}\BibitemShut {NoStop}%
\bibitem [{\citenamefont {Coretti}(2009)}]{CorettiThesis}%
  \BibitemOpen
  \bibfield  {author} {\bibinfo {author} {\bibfnamefont {S.}~\bibnamefont
  {Coretti}},\ }\href@noop {} {\enquote {\bibinfo {title} {Is (quantum)
  non-locality transitive?}}\ }\bibinfo {howpublished} {Semester Thesis, Swiss
  Federal Institute of Technology Z{\"u}rich} (\bibinfo {year}
  {2009})\BibitemShut {NoStop}%
\bibitem [{\citenamefont {Yin}(2021)}]{YinMasterThesis}%
  \BibitemOpen
  \bibfield  {author} {\bibinfo {author} {\bibfnamefont {Y.-C.}\ \bibnamefont
  {Yin}},\ }\emph {\bibinfo {title} {Transitivity of Quantum Nonlocality}},\
  \href {https://hdl.handle.net/11296/t7ahm2} {Master's thesis},\ \bibinfo
  {school} {National Cheng Kung University} (\bibinfo {year}
  {2021})\BibitemShut {NoStop}%
\bibitem [{\citenamefont {Cie{\'s}li{\'n}ski}\ \emph
  {et~al.}(2024)\citenamefont {Cie{\'s}li{\'n}ski}, \citenamefont {Knips},
  \citenamefont {Kowalczyk}, \citenamefont {Laskowski}, \citenamefont
  {Paterek}, \citenamefont {V{\'e}rtesi},\ and\ \citenamefont
  {Weinfurter}}]{CKK+24}%
  \BibitemOpen
  \bibfield  {author} {\bibinfo {author} {\bibfnamefont {P.}~\bibnamefont
  {Cie{\'s}li{\'n}ski}}, \bibinfo {author} {\bibfnamefont {L.}~\bibnamefont
  {Knips}}, \bibinfo {author} {\bibfnamefont {M.}~\bibnamefont {Kowalczyk}},
  \bibinfo {author} {\bibfnamefont {W.}~\bibnamefont {Laskowski}}, \bibinfo
  {author} {\bibfnamefont {T.}~\bibnamefont {Paterek}}, \bibinfo {author}
  {\bibfnamefont {T.}~\bibnamefont {V{\'e}rtesi}}, \ and\ \bibinfo {author}
  {\bibfnamefont {H.}~\bibnamefont {Weinfurter}},\ }\href {\doibase
  10.1073/pnas.2404455121} {\bibfield  {journal} {\bibinfo  {journal} {Proc.
  Natl. Acad. Sci. U.S.A.}\ }\textbf {\bibinfo {volume} {121}},\ \bibinfo
  {pages} {e2404455121} (\bibinfo {year} {2024})}\BibitemShut {NoStop}%
\bibitem [{\citenamefont {Christandl}\ \emph {et~al.}(2021)\citenamefont
  {Christandl}, \citenamefont {Ferrara},\ and\ \citenamefont
  {Horodecki}}]{CFH21}%
  \BibitemOpen
  \bibfield  {author} {\bibinfo {author} {\bibfnamefont {M.}~\bibnamefont
  {Christandl}}, \bibinfo {author} {\bibfnamefont {R.}~\bibnamefont {Ferrara}},
  \ and\ \bibinfo {author} {\bibfnamefont {K.}~\bibnamefont {Horodecki}},\
  }\href {\doibase 10.1103/PhysRevLett.126.160501} {\bibfield  {journal}
  {\bibinfo  {journal} {Phys. Rev. Lett.}\ }\textbf {\bibinfo {volume} {126}},\
  \bibinfo {pages} {160501} (\bibinfo {year} {2021})}\BibitemShut {NoStop}%
\bibitem [{\citenamefont {Farkas}\ \emph {et~al.}(2021)\citenamefont {Farkas},
  \citenamefont {Balanz\'o-Juand\'o}, \citenamefont {{\L}ukanowski},
  \citenamefont {Ko{\l}ody\'nski},\ and\ \citenamefont {Ac{\'\i}n}}]{FBL+21}%
  \BibitemOpen
  \bibfield  {author} {\bibinfo {author} {\bibfnamefont {M.}~\bibnamefont
  {Farkas}}, \bibinfo {author} {\bibfnamefont {M.}~\bibnamefont
  {Balanz\'o-Juand\'o}}, \bibinfo {author} {\bibfnamefont {K.}~\bibnamefont
  {{\L}ukanowski}}, \bibinfo {author} {\bibfnamefont {J.}~\bibnamefont
  {Ko{\l}ody\'nski}}, \ and\ \bibinfo {author} {\bibfnamefont {A.}~\bibnamefont
  {Ac{\'\i}n}},\ }\href {\doibase 10.1103/PhysRevLett.127.050503} {\bibfield
  {journal} {\bibinfo  {journal} {Phys. Rev. Lett.}\ }\textbf {\bibinfo
  {volume} {127}},\ \bibinfo {pages} {050503} (\bibinfo {year}
  {2021})}\BibitemShut {NoStop}%
\bibitem [{\citenamefont {Wiseman}\ \emph {et~al.}(2007)\citenamefont
  {Wiseman}, \citenamefont {Jones},\ and\ \citenamefont
  {Doherty}}]{WisemanPRL2007}%
  \BibitemOpen
  \bibfield  {author} {\bibinfo {author} {\bibfnamefont {H.~M.}\ \bibnamefont
  {Wiseman}}, \bibinfo {author} {\bibfnamefont {S.~J.}\ \bibnamefont {Jones}},
  \ and\ \bibinfo {author} {\bibfnamefont {A.~C.}\ \bibnamefont {Doherty}},\
  }\href {\doibase 10.1103/PhysRevLett.98.140402} {\bibfield  {journal}
  {\bibinfo  {journal} {Phys. Rev. Lett.}\ }\textbf {\bibinfo {volume} {98}},\
  \bibinfo {pages} {140402} (\bibinfo {year} {2007})}\BibitemShut {NoStop}%
\bibitem [{\citenamefont {Uola}\ \emph {et~al.}(2020)\citenamefont {Uola},
  \citenamefont {Costa}, \citenamefont {Nguyen},\ and\ \citenamefont
  {G\"uhne}}]{Steering-RMP}%
  \BibitemOpen
  \bibfield  {author} {\bibinfo {author} {\bibfnamefont {R.}~\bibnamefont
  {Uola}}, \bibinfo {author} {\bibfnamefont {A.~C.~S.}\ \bibnamefont {Costa}},
  \bibinfo {author} {\bibfnamefont {H.~C.}\ \bibnamefont {Nguyen}}, \ and\
  \bibinfo {author} {\bibfnamefont {O.}~\bibnamefont {G\"uhne}},\ }\href
  {\doibase 10.1103/RevModPhys.92.015001} {\bibfield  {journal} {\bibinfo
  {journal} {Rev. Mod. Phys.}\ }\textbf {\bibinfo {volume} {92}},\ \bibinfo
  {pages} {015001} (\bibinfo {year} {2020})}\BibitemShut {NoStop}%
\bibitem [{\citenamefont {Doherty}\ \emph {et~al.}(2002)\citenamefont
  {Doherty}, \citenamefont {Parrilo},\ and\ \citenamefont
  {Spedalieri}}]{DPS:PRL:2002}%
  \BibitemOpen
  \bibfield  {author} {\bibinfo {author} {\bibfnamefont {A.~C.}\ \bibnamefont
  {Doherty}}, \bibinfo {author} {\bibfnamefont {P.~A.}\ \bibnamefont
  {Parrilo}}, \ and\ \bibinfo {author} {\bibfnamefont {F.~M.}\ \bibnamefont
  {Spedalieri}},\ }\href {\doibase 10.1103/PhysRevLett.88.187904} {\bibfield
  {journal} {\bibinfo  {journal} {Phys. Rev. Lett.}\ }\textbf {\bibinfo
  {volume} {88}},\ \bibinfo {pages} {187904} (\bibinfo {year}
  {2002})}\BibitemShut {NoStop}%
\bibitem [{\citenamefont {Fang}\ \emph {et~al.}(2025)\citenamefont {Fang},
  \citenamefont {Tabia}, \citenamefont {Chen}, \citenamefont {Liang},\ and\
  \citenamefont {Lu}}]{FTC+25}%
  \BibitemOpen
  \bibfield  {author} {\bibinfo {author} {\bibfnamefont {X.-X.}\ \bibnamefont
  {Fang}}, \bibinfo {author} {\bibfnamefont {G.~N.~M.}\ \bibnamefont {Tabia}},
  \bibinfo {author} {\bibfnamefont {K.-S.}\ \bibnamefont {Chen}}, \bibinfo
  {author} {\bibfnamefont {Y.-C.}\ \bibnamefont {Liang}}, \ and\ \bibinfo
  {author} {\bibfnamefont {H.}~\bibnamefont {Lu}},\ }\href {\doibase
  10.1103/PhysRevLett.134.150201} {\bibfield  {journal} {\bibinfo  {journal}
  {Phys. Rev. Lett.}\ }\textbf {\bibinfo {volume} {134}},\ \bibinfo {pages}
  {150201} (\bibinfo {year} {2025})}\BibitemShut {NoStop}%
\bibitem [{\citenamefont {Navascu\'es}\ \emph {et~al.}(2007)\citenamefont
  {Navascu\'es}, \citenamefont {Pironio},\ and\ \citenamefont
  {Ac\'{\i}n}}]{NPA}%
  \BibitemOpen
  \bibfield  {author} {\bibinfo {author} {\bibfnamefont {M.}~\bibnamefont
  {Navascu\'es}}, \bibinfo {author} {\bibfnamefont {S.}~\bibnamefont
  {Pironio}}, \ and\ \bibinfo {author} {\bibfnamefont {A.}~\bibnamefont
  {Ac\'{\i}n}},\ }\href {\doibase 10.1103/PhysRevLett.98.010401} {\bibfield
  {journal} {\bibinfo  {journal} {Phys. Rev. Lett.}\ }\textbf {\bibinfo
  {volume} {98}},\ \bibinfo {pages} {010401} (\bibinfo {year}
  {2007})}\BibitemShut {NoStop}%
\bibitem [{\citenamefont {Navascu{\'e}s}\ \emph {et~al.}(2008)\citenamefont
  {Navascu{\'e}s}, \citenamefont {Pironio},\ and\ \citenamefont
  {Ac{\'\i}n}}]{NPA2008}%
  \BibitemOpen
  \bibfield  {author} {\bibinfo {author} {\bibfnamefont {M.}~\bibnamefont
  {Navascu{\'e}s}}, \bibinfo {author} {\bibfnamefont {S.}~\bibnamefont
  {Pironio}}, \ and\ \bibinfo {author} {\bibfnamefont {A.}~\bibnamefont
  {Ac{\'\i}n}},\ }\href {http://stacks.iop.org/1367-2630/10/i=7/a=073013}
  {\bibfield  {journal} {\bibinfo  {journal} {New J. Phys.}\ }\textbf {\bibinfo
  {volume} {10}},\ \bibinfo {pages} {073013} (\bibinfo {year}
  {2008})}\BibitemShut {NoStop}%
\bibitem [{\citenamefont {Chen}\ \emph {et~al.}()\citenamefont {Chen} \emph
  {et~al.}}]{KSChen_unpublished}%
  \BibitemOpen
  \bibfield  {author} {\bibinfo {author} {\bibfnamefont {K.-S.}\ \bibnamefont
  {Chen}} \emph {et~al.},\ }\href@noop {} {}\bibinfo {note} {(in
  preparation)}\BibitemShut {NoStop}%
\bibitem [{\citenamefont {Hsieh}\ \emph {et~al.}(2024)\citenamefont {Hsieh},
  \citenamefont {Tabia}, \citenamefont {Yin},\ and\ \citenamefont
  {Liang}}]{Hsieh2024resourcemarginal}%
  \BibitemOpen
  \bibfield  {author} {\bibinfo {author} {\bibfnamefont {C.-Y.}\ \bibnamefont
  {Hsieh}}, \bibinfo {author} {\bibfnamefont {G.~N.~M.}\ \bibnamefont {Tabia}},
  \bibinfo {author} {\bibfnamefont {Y.-C.}\ \bibnamefont {Yin}}, \ and\
  \bibinfo {author} {\bibfnamefont {Y.-C.}\ \bibnamefont {Liang}},\ }\href
  {\doibase 10.22331/q-2024-05-22-1353} {\bibfield  {journal} {\bibinfo
  {journal} {{Quantum}}\ }\textbf {\bibinfo {volume} {8}},\ \bibinfo {pages}
  {1353} (\bibinfo {year} {2024})}\BibitemShut {NoStop}%
\bibitem [{\citenamefont {Bowles}\ \emph {et~al.}(2014)\citenamefont {Bowles},
  \citenamefont {V\'ertesi}, \citenamefont {Quintino},\ and\ \citenamefont
  {Brunner}}]{BVQ+14}%
  \BibitemOpen
  \bibfield  {author} {\bibinfo {author} {\bibfnamefont {J.}~\bibnamefont
  {Bowles}}, \bibinfo {author} {\bibfnamefont {T.}~\bibnamefont {V\'ertesi}},
  \bibinfo {author} {\bibfnamefont {M.~T.}\ \bibnamefont {Quintino}}, \ and\
  \bibinfo {author} {\bibfnamefont {N.}~\bibnamefont {Brunner}},\ }\href
  {\doibase 10.1103/PhysRevLett.112.200402} {\bibfield  {journal} {\bibinfo
  {journal} {Phys. Rev. Lett.}\ }\textbf {\bibinfo {volume} {112}},\ \bibinfo
  {pages} {200402} (\bibinfo {year} {2014})}\BibitemShut {NoStop}%
\bibitem [{\citenamefont {Arora}\ and\ \citenamefont
  {Barak}(2009)}]{AroraBoaz2009book}%
  \BibitemOpen
  \bibfield  {author} {\bibinfo {author} {\bibfnamefont {S.}~\bibnamefont
  {Arora}}\ and\ \bibinfo {author} {\bibfnamefont {B.}~\bibnamefont {Barak}},\
  }\href@noop {} {\emph {\bibinfo {title} {Computational complexity: a modern
  approach}}}\ (\bibinfo  {publisher} {Cambridge University Press},\ \bibinfo
  {year} {2009})\BibitemShut {NoStop}%
\bibitem [{\citenamefont {Buhrman}\ \emph {et~al.}(2012)\citenamefont
  {Buhrman}, \citenamefont {Regev}, \citenamefont {Scarpa},\ and\ \citenamefont
  {Wolf}}]{Buhrman2012}%
  \BibitemOpen
  \bibfield  {author} {\bibinfo {author} {\bibfnamefont {H.}~\bibnamefont
  {Buhrman}}, \bibinfo {author} {\bibfnamefont {O.}~\bibnamefont {Regev}},
  \bibinfo {author} {\bibfnamefont {G.}~\bibnamefont {Scarpa}}, \ and\ \bibinfo
  {author} {\bibfnamefont {R.~d.}\ \bibnamefont {Wolf}},\ }\href {\doibase
  10.4086/toc.2012.v008a027} {\bibfield  {journal} {\bibinfo  {journal} {Theory
  Comput.}\ }\textbf {\bibinfo {volume} {8}},\ \bibinfo {pages} {623} (\bibinfo
  {year} {2012})}\BibitemShut {NoStop}%
\bibitem [{\citenamefont {Junge}\ and\ \citenamefont
  {Palazuelos}(2011)}]{Junge2011}%
  \BibitemOpen
  \bibfield  {author} {\bibinfo {author} {\bibfnamefont {M.}~\bibnamefont
  {Junge}}\ and\ \bibinfo {author} {\bibfnamefont {C.}~\bibnamefont
  {Palazuelos}},\ }\href {\doibase 10.1007/s00220-011-1296-8} {\bibfield
  {journal} {\bibinfo  {journal} {Commun. Math. Phys.}\ }\textbf {\bibinfo
  {volume} {306}},\ \bibinfo {pages} {695} (\bibinfo {year}
  {2011})}\BibitemShut {NoStop}%
\bibitem [{\citenamefont {Horodecki}\ \emph {et~al.}(1999)\citenamefont
  {Horodecki}, \citenamefont {Horodecki},\ and\ \citenamefont
  {Horodecki}}]{HHH99Teleportation}%
  \BibitemOpen
  \bibfield  {author} {\bibinfo {author} {\bibfnamefont {M.}~\bibnamefont
  {Horodecki}}, \bibinfo {author} {\bibfnamefont {P.}~\bibnamefont
  {Horodecki}}, \ and\ \bibinfo {author} {\bibfnamefont {R.}~\bibnamefont
  {Horodecki}},\ }\href {\doibase 10.1103/PhysRevA.60.1888} {\bibfield
  {journal} {\bibinfo  {journal} {Phys. Rev. A}\ }\textbf {\bibinfo {volume}
  {60}},\ \bibinfo {pages} {1888} (\bibinfo {year} {1999})}\BibitemShut
  {NoStop}%
\bibitem [{\citenamefont {Quintino}\ \emph {et~al.}(2016)\citenamefont
  {Quintino}, \citenamefont {Brunner},\ and\ \citenamefont
  {Huber}}]{Quintino2016}%
  \BibitemOpen
  \bibfield  {author} {\bibinfo {author} {\bibfnamefont {M.~T.}\ \bibnamefont
  {Quintino}}, \bibinfo {author} {\bibfnamefont {N.}~\bibnamefont {Brunner}}, \
  and\ \bibinfo {author} {\bibfnamefont {M.}~\bibnamefont {Huber}},\ }\href
  {\doibase 10.1103/PhysRevA.94.062123} {\bibfield  {journal} {\bibinfo
  {journal} {Phys. Rev. A}\ }\textbf {\bibinfo {volume} {94}},\ \bibinfo
  {pages} {062123} (\bibinfo {year} {2016})}\BibitemShut {NoStop}%
\bibitem [{\citenamefont {Jones}\ \emph {et~al.}(2007)\citenamefont {Jones},
  \citenamefont {Wiseman},\ and\ \citenamefont {Doherty}}]{JWD:PRA:2007}%
  \BibitemOpen
  \bibfield  {author} {\bibinfo {author} {\bibfnamefont {S.~J.}\ \bibnamefont
  {Jones}}, \bibinfo {author} {\bibfnamefont {H.~M.}\ \bibnamefont {Wiseman}},
  \ and\ \bibinfo {author} {\bibfnamefont {A.~C.}\ \bibnamefont {Doherty}},\
  }\href {\doibase 10.1103/PhysRevA.76.052116} {\bibfield  {journal} {\bibinfo
  {journal} {Phys. Rev. A}\ }\textbf {\bibinfo {volume} {76}},\ \bibinfo
  {pages} {052116} (\bibinfo {year} {2007})}\BibitemShut {NoStop}%
\bibitem [{\citenamefont {Zhang}\ and\ \citenamefont
  {Chitambar}(2024)}]{zhang2024exact}%
  \BibitemOpen
  \bibfield  {author} {\bibinfo {author} {\bibfnamefont {Y.}~\bibnamefont
  {Zhang}}\ and\ \bibinfo {author} {\bibfnamefont {E.}~\bibnamefont
  {Chitambar}},\ }\href {\doibase 10.1103/PhysRevLett.132.250201} {\bibfield
  {journal} {\bibinfo  {journal} {Phys. Rev. Lett.}\ }\textbf {\bibinfo
  {volume} {132}},\ \bibinfo {pages} {250201} (\bibinfo {year}
  {2024})}\BibitemShut {NoStop}%
\bibitem [{\citenamefont {Renner}(2024)}]{Renner2024PRL}%
  \BibitemOpen
  \bibfield  {author} {\bibinfo {author} {\bibfnamefont {M.~J.}\ \bibnamefont
  {Renner}},\ }\href {\doibase 10.1103/PhysRevLett.132.250202} {\bibfield
  {journal} {\bibinfo  {journal} {Phys. Rev. Lett.}\ }\textbf {\bibinfo
  {volume} {132}},\ \bibinfo {pages} {250202} (\bibinfo {year}
  {2024})}\BibitemShut {NoStop}%
\end{thebibliography}
\end{document}